%% file: balmvlt.tex
\documentclass[usenatbib]{mn2e}

\usepackage{graphicx}

\newcommand{\lammax}{\lambda_{\rm max}}
\newcommand{\pmax}{P_{\rm max}}
\newcommand{\thetapa}{\theta_{\rm PA}}
\newcommand{\ebv}{E_{B-V}}
\newcommand{\flam}{F_{\lambda}}

\title[The Balmer Edge of Quasars]{
The buried Balmer-edge signatures from quasars\footnotemark[2]}

\author[Kishimoto, Antonucci, Boisson \& Blaes]{
Makoto Kishimoto$^{1}$\thanks{E-mail: mk@roe.ac.uk},
Robert Antonucci$^{2}$,
Catherine Boisson$^{3}$
and Omer Blaes$^{2}$ \\
$^{1}$Institute for Astronomy, University of
Edinburgh, Blackford Hill, Edinburgh EH9 3HJ, UK\\ 
$^{2}$Physics Department, 
University of California, Santa Barbara, CA 93106, USA\\
$^{3}$LUTH, FRE 2462 du CNRS, associ\'ee \`a l'Universit\'e Denis Diderot,
    Observatoire de Paris, Section de Meudon, \\ 
    F--92195 Meudon Cedex, France}

\begin{document}

\date{Accepted for publication in MNRAS}

\pagerange{\pageref{firstpage}--\pageref{lastpage}} \pubyear{2004}

\maketitle

\label{firstpage}

\begin{abstract}

In our previous paper, we have reported the detection of a Balmer edge
absorption feature in the polarized flux of one quasar (Ton 202).  We
have now found similar Balmer edge features in the polarized flux of four
more quasars (4C09.72, 3C95, B2 1208+32, 3C323.1), and possibly a few
more, out of 14 newly observed with the VLT and Keck telescopes.  In
addition, we also re-observed Ton 202, but we did not detect such a
dramatic feature, apparently due to polarization variability (the two
observations are one-year apart).  The polarization measurements of some
quasars are affected by an interstellar polarization in our Galaxy, but
the measurements have been corrected for this effect reasonably well.

Since the broad emission lines are essentially unpolarized and the
polarization is confined only to the continuum in the five quasars
including Ton 202 in both epochs, the polarized flux is considered to
originate interior to the broad emission line region. The Balmer edge
feature seen in the polarized flux is most simply interpreted as an
intrinsic spectral feature of the quasar UV/optical continuum, or the
``Big Blue Bump'' emission. In this case, the edge feature seen in
absorption indeed indicates the thermal and optically-thick nature
of the continuum emitted.  However, we also discuss other possible 
interpretations.

\end{abstract}


\footnotetext[2]{Partially based on observations collected at the
European Southern Observatory,Chile (ESO Programme 69.B-0566A)}

\section{Introduction}\label{sec-intro}

We have recently reported our first detection of a Balmer edge absorption
feature in the polarized flux spectrum of one quasar (Ton 202;
\citealt*{KAB03}; hereafter Paper I). In the present paper, we report
that a similar Balmer edge feature is seen in the relatively
high S/N polarized flux spectrum of four more quasars, taken with the
Very Large Telescope (VLT) and Keck Telescope.  We also present new
spectropolarimetric data for 10 more quasars of a mostly moderate S/N,
also taken with the VLT and Keck.

These observations are a part of our on-going effort to find spectral
features intrinsic to the UV/optical continuum of quasars, which is
often called the Big Blue Bump (BBB).  This emission component is
crucial in the sense that it dominates the radiative output of quasars,
but its nature has not been well understood \citep{KB99,An88,An99}.  One
of the most critical problems has been the observed apparent lack of 
continuum edge spectral features from the emission source,
which is often assumed to be an accretion disk.

At least at the Lyman edge, this problem may be mitigated by smearing from 
relativistic Doppler shifts and gravitational redshifts, which can 
severely blur the spectral features produced in the local atmospheres at 
different radii of the disk (e.g. Hubeny et al.  2000).  While the coolest 
disk models still exhibit a broadened absorption feature, hotter disk 
models generally exhibit a change in slope in the vicinity of the Lyman 
limit.  This arises from a combination of smeared out absorption and 
emission edges on top of a general rollover of the underlying continuum. 
Such slope changes might be consistent with the spectral breaks that are 
observed in the vicinity of the Lyman limit in some objects such as 3C~273 
(Kriss et al. 1999) or in composite spectra (Zheng et al. 1997; Telfer et 
al. 2002), but quantitative fits to such spectra have so far been 
unsatisfactory.  In particular, detailed fits to the spectral break 
observed in 3C~273 still produce a local emission bump that is not in 
agreement with the data (Blaes et al. 2001).  This bump in the best fit 
disk model arises from Lyman emission edges formed in the atmospheres 
between ten and thirty gravitational radii.  Relativistic smearing of this 
bump is minimized because 3C~273 is a superluminal source and the putative 
disk is therefore constrained to be nearly face-on to the line of sight. 
More generally, disk models that provide qualitative fits to the Lyman 
spectral break still do not provide enough emission in the far ultraviolet 
and soft X-rays to fit the composite spectra because the thermal continuum 
rolls over too soon (Laor et al. 1997; Blaes 2004).  Some additional 
source of emission, e.g. Comptonization (Czerny \& Elvis 1987; Lee, Kriss 
\& Davidsen 1992), is probably required.

Compared to the Lyman edge region, accretion disk models place the main
emitting source for the Balmer edge region considerably farther out in
the gravitational potential well.  In detail, such models predict
substantial contributions to both the Lyman and Balmer edge regions from
broad ranges of overlapping radii, but the Balmer edge still has contributions 
from considerably larger radii, and therefore will be subject to less
smearing.  (There are actually considerable technical problems in making
concrete spectral predictions of the Balmer edge from accretion disk
models because of the presence of hydrogen ionization zones at these
larger radii - see \citealt{Hu00} for discussion.  It is noteworthy,
however, that the Balmer edges in these models is always in absorption.)
In addition, unlike the Lyman edge, the Balmer edge is not a resonant
feature, so the confusion with foreground absorption is much less of a
problem.  Thus, although detailed spectral predictions from disk
atmosphere models of the Balmer edge are yet to come, targeting the
Balmer edge may well be advantageous compared to the Lyman edge.

However, the observation of the Balmer edge intrinsic to the BBB
emission has been impossible, since it is essentially buried under the
Balmer continuum, high-order Balmer emission lines, and FeII emission
lines (collectively referred to as the small blue bump) which are
thought to be coming from the broad emission line region (BLR).  The
main immediate goal of our effort is to overcome this obstacle by
looking at the polarized flux of the quasars with a small polarization
($P\sim1$\%).  Note that these are normal quasars in terms of their
spectral energy distribution, and not classified as blazars. About 1/3
of the normal quasars are known to be polarized with $P > 0.6$\%
\citep{Be90}.  Among them, at least in some cases (and possibly in many
cases), the polarization is found to be confined only to the continuum,
and the broad emission lines are essentially unpolarized
(\citealt{An88}, showing the data of Miller and Goodrich;
\citealt{SS00}). In these cases, the polarized flux is considered to
arise {\it interior to} the BLR, and thus it is likely to show the
intrinsic Balmer edge feature of the BBB emission with all contaminating
emission from outside the nucleus scraped off and removed.

Detailed investigation of the polarized flux requires a long
integration with a large telescope.  In our first such deep integration
on one quasar with the Keck telescope, we have indeed found a Balmer
edge feature in the polarized flux of the quasar. Here, we report our
spectropolarimetric observations of 14 more quasars with the VLT and
Keck, presenting a similar Balmer edge feature in four more quasars. We
also discuss the re-observation of Ton202. Some of the preliminary
results have been published in Kishimoto et al. (2004). 

Finally, note that the polarization $P$ and the polarized flux $P \times
F$ are different quantities which must not be confused.  It has been
previously established that $P$ tends to be low in the Balmer continuum
region, but our work, which has important new physical implications, is
that $P \times F$ in the same wavelength region is also well below the
extrapolation from longer wavelengths in many cases.

For convenience, we summarize the symbols and acronyms used throughout
this paper in Table~\ref{tab-symbol}.

\begin{table}
  \caption{List of symbols and acronyms used in this paper.}
  \begin{tabular}{p{1.5cm}p{6cm}}
  \hline
  symbol/ acronym        & definition \\

  \hline

  $I$ & Stokes $I$ parameter, or the total flux $F$. \\

  $Q, U$ & Unnormalized Stokes parameters in the sky
   coordinates,   i.e. having the reference axis along the North-South direction. \\

  $q, u$ & Normalized Stokes parameters, $q=Q/I$, $u=U/I$. \\

  $Q', U', q', u'$ & Rotated Stokes parameters 
            with the reference axis 
            taken to be along the PA 
            of the polarization 
            of the object. Thus, $Q'$ (or $q' \times F$) is essentially
   equivalent to the polarized flux $P \times F$.\\

  $r_P$ & Ratio of polarization at 2891-3600\AA\  
          to that at 4000-4731\AA.  \\ 

  $\Delta$PA & Difference between the PA of the optical 
               polarization and that of the radio structural axis. \\

  $\lammax$ & The wavelength at which an interstellar 
             polarization is maximum for a given line of sight.\\
  $\pmax$ & The magnitude of an interstellar polarization 
           at $\lammax$. \\

  \\ 

  PA & Position Angle. \\
  BLR & Broad Line Region. \\
  BBB & Big Blue Bump. \\
  ISP & Interstellar Polarization. \\

  \hline
  \end{tabular}
  \label{tab-symbol}
\end{table}

\begin{table*}
\begin{minipage}{110mm}
  \caption{Log of VLT/FORS1 observations in 2002. The slit PA was fixed at 0\degr.}
  \begin{tabular}{llccccc}
  \hline
  name        & other     & $z$   & exp.          & date        & slit   &   filter \\
              & name      &       & (min)         &             & ($''$) & \\
  \hline

  Q0003+158   & 4C15.01   & 0.450 & 20 $\times$ 1 & 13 Sep & 1.5 & GG375 \\
              &           &       & 20 $\times$ 1 & 15 Sep & 1.5 & GG375 \\
  Q0205+024   &           & 0.155 & 10 $\times$ 1 & 13 Sep & 1.5 & none  \\
  Q0349-146   & 3C95      & 0.616 & 30 $\times$ 3 & 13 Sep & 1.5 & GG435 \\
              &           &       & 16 $\times$ 1 & 13 Sep & 1.5 & GG435 \\
              &           &       & 30 $\times$ 2 & 15 Sep & 1.5 & GG435 \\
              &           &       & 30 $\times$ 3 & 15 Sep & 1.5 & none  \\
              &           &       & 20 $\times$ 1 & 15 Sep & 1.5 & none  \\
  Q0350-073   & 3C94      & 0.962 & 20 $\times$ 1 & 13 Sep & 1.5 & GG435 \\      
  Q0405-123   & PKS       & 0.573 & 10 $\times$ 1 & 13 Sep & 1.0 & GG435 \\        
  Q0414-060   & 3C110     & 0.775 & 10 $\times$ 1 & 13 Sep & 1.5 & GG435 \\      
  Q1912-550   & PKS       & 0.402 & 20 $\times$ 1 & 13 Sep & 1.0 & GG375 \\
              &           &       & 20 $\times$ 5 & 15 Sep & 1.5 & GG375 \\
              &           &       & 30 $\times$ 1 & 15 Sep & 1.5 & GG375 \\
  Q2115-305   & PKS       & 0.979 & 10 $\times$ 1 & 13 Sep & 1.0 & GG435 \\
              &           &       & 10 $\times$ 3 & 15 Sep & 1.0 & GG435 \\
  Q2251+113   & 4C11.72   & 0.326 & 10 $\times$ 1 & 13 Sep & 1.0 & none  \\
  Q2308+098   & 4C09.72   & 0.433 & 30 $\times$ 2 & 13 Sep & 1.0 & GG375 \\
              &           &       & 30 $\times$ 2 & 13 Sep & 1.5 & GG375 \\
  Q2349-014   & 4C-01.61  & 0.174 & 10 $\times$ 1 & 13 Sep & 1.0 & none  \\

  \hline
  \end{tabular}
  \label{tab-log}
\end{minipage}
\end{table*}

\begin{table*}
\begin{minipage}{110mm}
  \caption{Log of Keck observations on 4 May 2003. The slit width was $1.''5$.}
  \begin{tabular}{llcccc}
  \hline
  name        & other       & $z$   & exp.        & slit PA & grism\\
              & name        &       & (min)       & (\degr) & \\
  \hline

  Q1004+130   & 4C13.41     & 0.240 & 30$\times$1 &  58 & 300/5000 \\ 
  Q1208+322   & B2 1208+32  & 0.388 & 30$\times$3 &  30 & 300/5000 \\
              &             &       & 30$\times$1 &  30 & 400/3400 \\
  Q1425+267   & Ton 202     & 0.366 & 30$\times$2 & 143 & 300/5000 \\
              &             &       & 30$\times$1 &  97 & 400/3400 \\
  Q1545+210   & 3C323.1     & 0.264 & 30$\times$2 &  85 & 300/5000 \\
              &             &       & 16$\times$1 &  85 & 400/3400 \\

  \hline
  \end{tabular}
  \label{tab-log-keck}
\end{minipage}
\end{table*}

\section{Observation}\label{sec-obs}

\subsection{VLT observations}

We observed 11 quasars on 13 and 15 September 2002 (UT) using Focal
Reducer/Low Dispersion Spectrograph (FORS1) on the Unit Telescope 3
(Melipal) of the Very Large Telescope (VLT). These observations are
summarized in Table \ref{tab-log}.  The observation of each target were
done in one or more sequences with one sequence consisting of four
frames with four different waveplate positions. The exposure time in
Table \ref{tab-log} refers to the exposure time for one sequence multiplied 
by the number of repetitions of the same sequence.  The grism 300V was used
at a dispersion of 2.6\AA\ per pixel.  The seeing varied between
$\sim0.''5$ and $\sim1''$ over the nights: the slit width was set to
$1.''5$ giving a spectral resolution of 20\AA, except for some frames
taken in the first half of 13 Sep (see Table \ref{tab-log}) when the
seeing was mostly $0.''5-0.''7$ and the slit width was set to $1''$
(resolution 13\AA).  The spatial sampling was $0.''200$ per pixel. The
slit direction was fixed north-south, and the instrument was equipped 
with an atmospheric dispersion compensator \citep{ARB97}.

The CCD frames were bias-subtracted using an averaged bias frame, and
flat-fielded using internal screen flats. Obvious cosmic-ray hits were
removed and interpolated using neighboring pixels. The o-ray and e-ray
spectra were extracted with $2.''2$ windows. Then these were combined to
produce normalized Stokes parameters $q$ and $u$, and the Stokes parameter
$I$, following \citet*{MRG88}.  The zero point of the position angle
(PA) of the polarization was determined using the polarized standard
stars Hiltner 652, BD+25 727, and BD-12-5133, with an estimated
accuracy of 0.3\degr.  The instrumental polarization was checked with
the unpolarized standard stars HD13588 and HD42078 and estimated to be
less than $\sim 0.05$\%. The flux was calibrated using the flux standard
star GD50.

In many of our observations, we have used two different order-sorting
filters to avoid second-order contamination; we have chosen for each
object the filter which blocks the second order spectrum at the H$\beta$
line position in the observed frame. We observed one of the objects
(3C95) also without the filter to cover the near-UV wavelengths.

\subsection{Keck observations}

Four quasars listed in Table~\ref{tab-log-keck} were observed on 4 May
2003 (UT) with the Low Resolution Imaging Spectrograph (LRIS;
\citealt{Ok95}) on the Keck-I telescope, with a new blue-sensitive CCD
(commissioned on 4 June 2002, after the observations in Paper I) on the
blue arm of the spectrograph.  Two grisms, 300/5000 and 400/3400, were
used with dispersions of 1.4\AA\ per pixel and 1.0\AA\ per pixel,
respectively.  The slit width was $1.''5$, giving a spectral
resolution of $\sim$14\AA\ with grism 300/5000 and $\sim$10\AA\ with
400/3400.  The spatial sampling along the slit was $0.''135$ per
pixel, and the data were obtained with the slit roughly at the
parallactic angle.  The seeing size was estimated to be about $0.''8$
in the first half of the night. We probably were affected by some
cirrus especially in the second half of the night.

The 2D spectral images were reduced in the standard manner. The bias
level was subtracted using bias frames and overscan regions.  The o-ray
and e-ray spectra were extracted with $2.''8$ window. Then these spectra
taken at 4 positions of the waveplate were combined following
\citet{MRG88}.  The zeropoint of the PA was calibrated with the
polarized standard star HD183143.  The instrumental polarization was
checked with the unpolarized standard stars HD94851 and BD+28 4211 and
also with the observations of Ton202 with two different slit PAs. The
former two observations indicated possible instrumental polarization of
$P \sim 0.13$\% roughly independent of wavelengths at a certain PA in the
instrumental coordinates.  However, the latter observations of Ton202 at
two PAs showed no systematic difference (the correction with 0.13\%
results in systematic difference of this level).  Therefore, we did not
correct for any instrumental polarization.  The flux was calibrated
using the standard stars Feige 34 and BD+28 4211.

For the observations with the grism 300/5000, there was no order sorting
filter available, and the new blue arm CCD is very blue
sensitive. Therefore, the spectra at the red side are contaminated by
second-order light.  To correct for the contamination in flux, we
implemented short exposures on Ton202 and B2 1208+32 with dichroic D460,
which cuts out the light with wavelengths longer than
$\sim4600$\AA. This provided the count rate spectrum for the second
order light (the spectrum $\la 5500$\AA\ was masked). We scaled and
subtracted it from the Stokes $I$ spectra of these two
objects. Therefore, the total flux spectrum is expected to contain very
little contamination. However, the polarization measurement was not
corrected (the observations with D460 were done only at one position
angle of the waveplate), so the polarized flux spectrum could be
slightly affected.  For 3C323.1 and Q1004+130, the spectra were not
corrected, but these have lower redshift than the other two objects, so
the wavelength range up to H$\beta$ line is probably not affected
severely.

\subsection{Targets}

The selected quasars were known to be polarized in a broad optical band
\citep{SMA84,Be90,VW98}. We observed the ones with redshifts larger than
$\sim0.2$ so that the Balmer edge in the observed frame is well away from
the atmospheric cut off and well within the detector sensitivity
range. Most of them are at redshift less than $\sim$0.6 in order to cover
the H$\beta$ line. In the VLT observations, we also observed a few higher
redshift objects to explore the Balmer continuum under the small blue bump.
Four quasars in our sample (3C95, 4C09.72, B2 1208+32, and 3C323.1) had
already been observed spectropolarimetrically \citep{SS00} and had been
found to have unpolarized broad lines with limited S/N. We have carried
out long integrations on them to scrutinize the polarization of the broad
emission lines and the Balmer edge wavelengths. Ton 202 had been
observed spectropolarimetrically by \citet{SS00} and by us (Paper I),
and was included in our sample for re-observation.  In the VLT run,
other quasars were observed with a short integration in the first
night, and we tried to follow up in the second night the ones which
seemed to show unpolarized broad lines and/or unpolarized small blue
bump (see next section).

\subsection{Interstellar polarization}

Interstellar polarization (ISP) can affect our observations.  For
stars with $\ebv \ga 0.1$ mag, \citet*{SMF75} showed that $\pmax \la
9\ebv$ where $\pmax$ is the maximum polarization in percent over the
(optical) wavelengths.  We list $\ebv$ and $9\ebv$ as well as the
Galactic coordinates $(l,b)$ in Table~\ref{tab-isp}.  However, the
limit of $9\ebv$ are less certain for the lines of sight with $\ebv
\la 0.1$ mag (see Figure 9 of \citealt{SMF75}).  A better constraint
can be obtained from the polarization measurements of Galactic stars
in each quasar field.  We have checked the polarization catalog of
Galactic stars by \citet{He00}, and have tried to observe a star for
each quasar with a suspected ISP effect and/or for quasars observed with a
long integration time. The stars were chosen to be nearby in the
projected distance to each quasar but far enough in the real distance
(essentially a faint early-type star or a faint giant) in order to
sample the whole ISP in our Galaxy along the line of sight.  We
obtained the measurement for four quasars (4C09.72, 3C95, Q1912-550,
3C323.1), as summarized in Table \ref{tab-log-star}.

\begin{table}
  \caption{Properties related to interstellar polarization.} 
  \label{tab-isp}
  \begin{tabular}{lcccccccl}
  \hline
  Name         & $\ebv$ & 9$\ebv$ & $(l,b) (\degr)$ \\ 
  \hline      
	      
  Q0003+158    & 0.050 & 0.45 & (107,-45) \\ 
  Q0205+024    & 0.029 & 0.26 & (158,-55) \\ 
  3C95         & 0.079 & 0.71 & (205,-46) \\ 
  Q0350-073    & 0.079 & 0.71 & (197,-43) \\ 
  Q0405-123    & 0.058 & 0.52 & (205,-42) \\ 
  Q0414-060    & 0.043 & 0.39 & (199,-37) \\ 
  Q1004+130    & 0.038 & 0.34 & (225,+49) \\ 
  B2 1208+32   & 0.017 & 0.15 & (182,+80) \\ 
  Ton 202      & 0.019 & 0.17 & (36, +68) \\ 
  3C323.1      & 0.042 & 0.38 & (34, +49) \\ 
  Q1912-550    & 0.059 & 0.53 & (342,-25) \\ 
  Q2115-305    & 0.120 & 1.1  & ( 16,-44) \\ 
  Q2251+113    & 0.086 & 0.77 & ( 83,-42) \\ 
  4C09.72      & 0.042 & 0.38 & ( 86,-46) \\ 
  Q2349-014    & 0.027 & 0.24 & ( 92,-60) \\ 

  \hline
  \end{tabular}

  \medskip

  $\ebv$ is from NED (data from \citealt{SFD98}).  $9\ebv$ is a
  maximum expected interstellar polarization along the line of sight to
  each quasar, but relatively uncertain for these low $\ebv$ cases.
  $(l,b)$ is Galactic coordinates.

\end{table}

\begin{table*}
\begin{minipage}{150mm}

  \caption{The interstellar polarization measurements of Galactic
  stars around some of our quasars.  Distances are from spectral type
  classifications, except for SAO245987 where it is from the measured
  trigonometric parallax. $\lammax$ is the wavelength at which the
  polarization is maximum and $\pmax$ is that maximum polarization at
  $\lammax$.}

  \begin{tabular}{llcccc}
  \hline
  Galactic star & projected distance          & distance (kpc)  & $\pmax$ (\%)      & $\lammax$ (\AA) & PA(V) (\degr)   \\
										                    		      
  \hline									                    		      
										                    		      
  HD23938       & 1.2\degr\ NW from 3C95      & $\sim 0.57$     & $<0.1$            & -              & -                \\
  AG+22 1545    & 1.3\degr\ N from 3C323.1    & $\sim 1.3 $     & $0.718 \pm 0.005$ & $5570 \pm 130$ & $84.3  \pm 0.4$  \\
  SAO245987     & 0.5\degr\ SW from Q1912-550 & $0.24 \pm 0.08$ & $0.715 \pm 0.005$ & $5530 \pm 110$ & $  1.9 \pm 0.3$  \\
  Pn23s1-15     & 0.4\degr\ NW from 4C09.72   & $\sim 3.3$      & $0.305 \pm 0.007$ & $4970 \pm 420$ & $104.7 \pm 1.3$  \\

  \hline
  \end{tabular}
  \label{tab-log-star}
\end{minipage}
\end{table*}

\begin{table*}
\begin{minipage}{150mm}
  \caption{Broadband polarization properties.  The measurements before any ISP
  correction are shown.  For Q0003+158, 3C95, 3C323.1, Q1912-550, and 4C09.72,
  an additional row right below the row for each object presents 
  the measurements after the ISP correction, with the first column
  showing the amount of ISP adopted.}
  \label{tab-pol}
  \input{tab_pol.tex}
  \\
  \medskip
  $^*$The absolute PA for the May 2002 data of Ton 202 was not determined and  
only the relative PA was measured.

\end{minipage}                                                              
\end{table*}

\begin{table*}
\begin{minipage}{190mm}
  \caption{H$\beta$ and [OIII] line properties. See the caption of
  Table~\ref{tab-pol} for the ISP correction adopted.}
  \label{tab-hbeta}
  \input{tab_hb.tex}
  \\
  \medskip
  $^*$The absolute PA for the May 2002 data of Ton 202 was not determined and  
only the relative PA was measured.

\end{minipage}
\end{table*}

\begin{table*}
\begin{minipage}{150mm}
  \caption{MgII line properties. Continuum region is
 2611-2705\AA\ and 2891-2985\AA. 
 Line region is 2705-2891\AA. See the caption of
  Table~\ref{tab-pol} for the ISP correction adopted.}
  \label{tab-mg}
  \input{tab_mg.tex}
  \\
  \medskip
  $^*$The absolute PA for the May 2002 data of Ton 202 was not determined and  
only the relative PA was measured.

\end{minipage}
\end{table*}

\section{Results}

Our primary targets are the quasars with a small continuum-confined
polarization, showing unpolarized broad emission lines. In these
quasars, we expect in principle the small blue bump also to be
unpolarized.  Thus, these quasars would show a polarization which
drops significantly at wavelengths shortward of 4000\AA\ in the rest
frame. This is a broad band property and relatively easy to recognize.
We have measured this polarization drop as the ratio $r_P$ of the
polarization at 2891-3600\AA\ to that at 4000-4731\AA. Those are
listed in Table~\ref{tab-pol} (polarization in this table has been
debiased following \citealt{SS85}).

We also have measured the polarization of the emission lines by fitting
a power-law in the adjcent continuum region and subtracting off the
continuum both in the total flux and polarized flux. More precisely, we
first measured the PA of the continuum polarization, and rotated the
Stokes parameters to that PA, and fit a power-law and subtracted it off.
Then we measured the PA of the residual Stokes spectra, rotate the
Stokes parameters to that PA, and measured the flux.  Then we calculated
the equivalent width (EW) of the polarized line flux using the fitted
continuum (although the PA could be different). The results are shown in
Table~\ref{tab-hbeta} and Table~\ref{tab-mg} (quoted polarization in
these two tables is the normalized, rotated Stokes $q$ with the
reference axis along the quoted PA, and thus has not been debiased).

For the H$\beta$ line, we measured the polarization and EW of the
H$\beta$ line at $4731-4934$\AA, i.e. the wavelength region of
[OIII]$\lambda$4959,5007 lines were excluded to avoid the
effect of these lines on the H$\beta$ measurements. The EW thus
derived is smaller than the real EW. This is not a problem for our
primary purpose here, which is to determine whether or not the line is
detected in the polarized flux and to determine its polarization if
the line is detected.  The EW of the line in total flux was measured
using the same wavelength window, to compare directly with the EW in
the polarized flux. The measurements for [OIII]$\lambda$5007 line and
MgII$\lambda$2798 line were done at 4982-5032\AA\ and 2705-2891\AA,
respectively.

Note that the absorption features at the observed wavelenths of $\sim
6900$\AA\ and $\sim 7600$\AA\ seen in the total flux and polarized flux
spectra are telluric.  These regions were excluded from
the measurements (thus, also due to this exclusion, the EWs measured
here can be smaller than the true EWs).

In the following, we first describe the results for the quasars where
the ratio $r_P$ is measured to be significantly less than unity with
$\sim 3 \sigma$.  The majority of the objects in our sample show this
significant drop in $P$: 4C09.72, 3C95, 3C323.1, B2 1208+32, Ton202,
Q1912-550, Q0003+158, Q0405-123, Q2349-014.  Then we briefly describe
the results for the other six quasars: for these, we did not detect
a significant $P$ decrease shortward of 4000\AA, although the constraint
is not tight in some of the objects.  The polarizations of Galactic
stars around each target other than those listed in
Table~\ref{tab-log-star} are based on the compilation by \citet{He00}.

All flux has been corrected for the Galactic reddening using the values
of $\ebv$ in Table~\ref{tab-isp} and the reddening curve of
\citet*{Ca89} with $R_V=3.1$.

\begin{figure}
 \includegraphics[width=80mm]{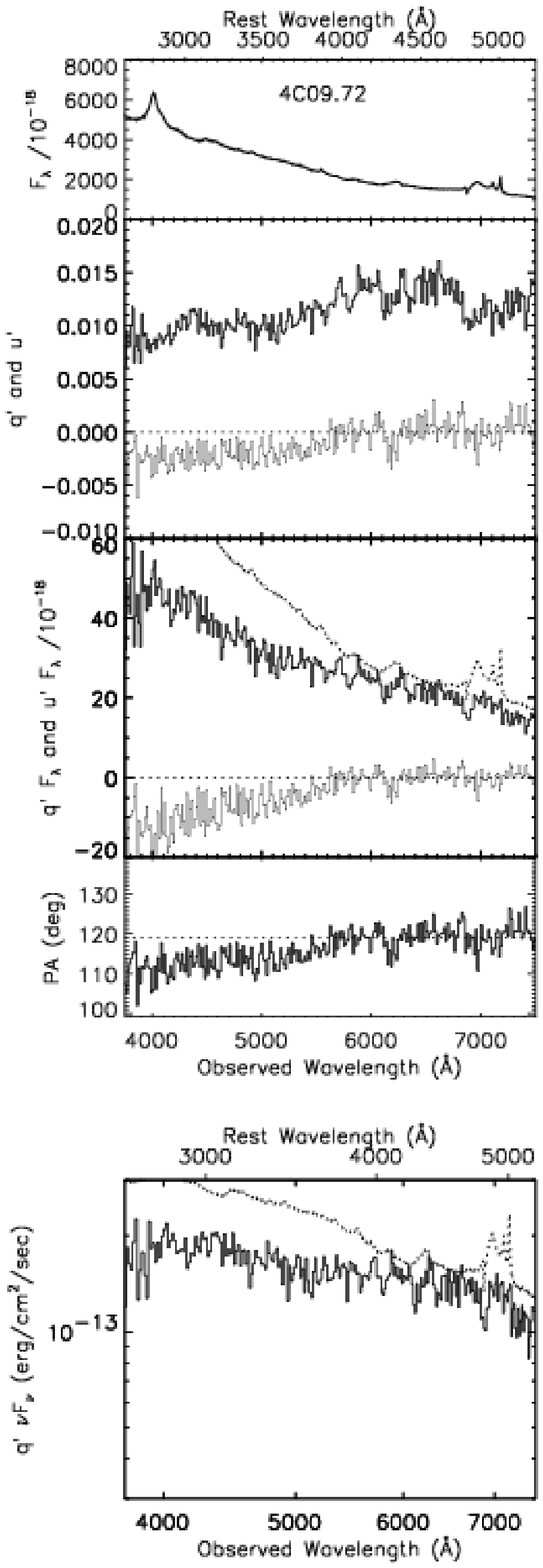} 
 \caption{ VLT/FORS1 spectropolarimetric measurements for 4C09.72. From
 top to bottom, Stokes $I$, 
 normalized $q'$ and $u'$, unnormalized $Q'$
 and $U'$ (the referece axis is along the polarization PA at
 4000-4731\AA) with scaled $I$ (dotted line), and PA. 
 The fluxes are in units of 10$^{-18}$ erg cm$^{-2}$ sec$^{-1}$ \AA$^{-1}$.
 The separate bottom panel
 is a log-log plot of $q' \times \nu F_{\nu}$ and $\nu F_{\nu}$
 (dotted line) against $\lambda$ (note the different scale in
 $x$-axis).}
 \label{res-4C09.72}
\end{figure}

\begin{figure}
 \includegraphics[width=80mm]{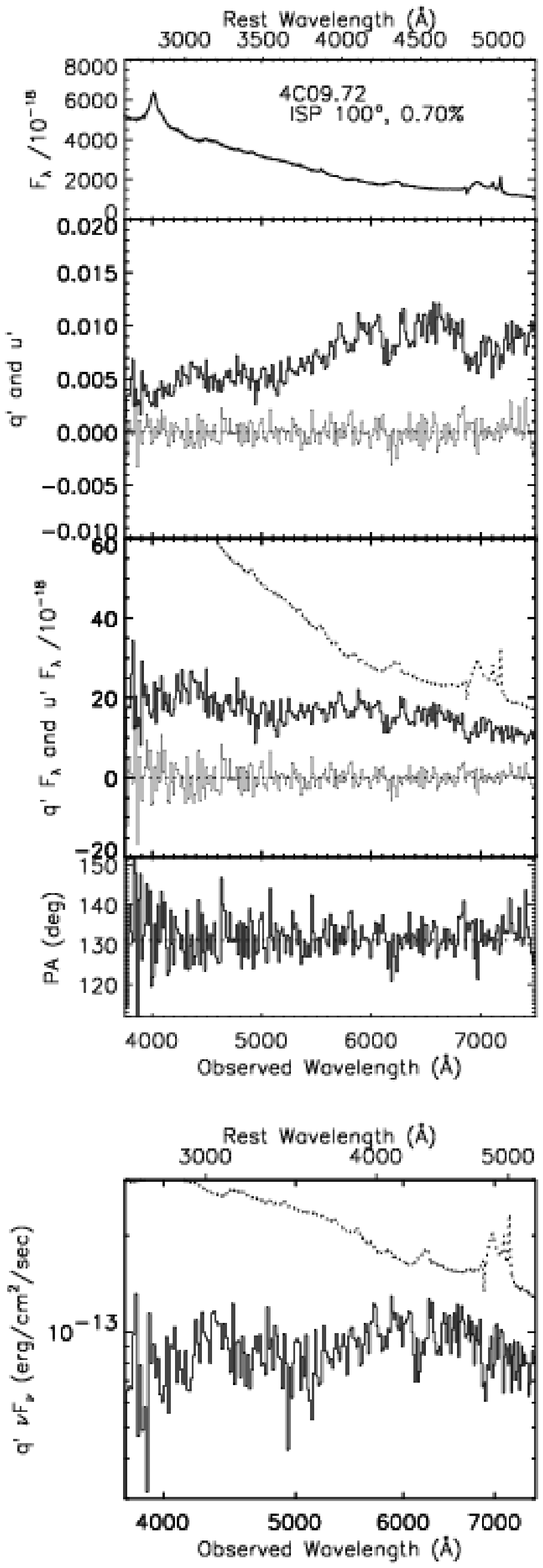}
 \caption{The ISP-corrected polarization of 4C09.72, assuming ISP PA
 of 100\degr.}
 \label{res-4C09.72-cor100}
\end{figure}

\begin{figure}
 \includegraphics[width=80mm]{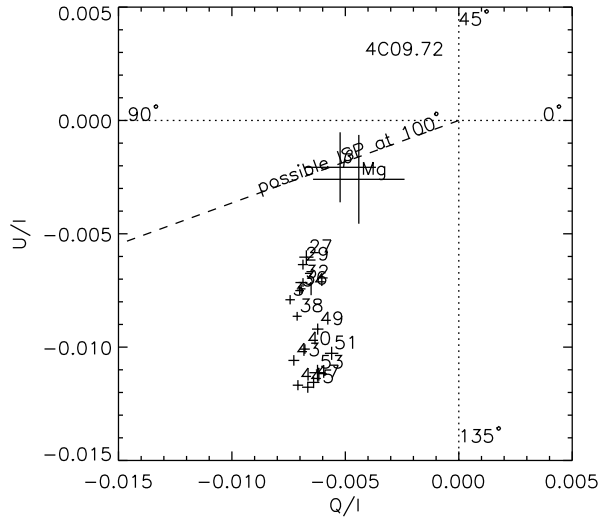}
 \caption{The data on the plane of normalized Stokes parameters $q$
 and $u$ with error bars. The numbers beside each data
 point are wavelengths in units of 100\AA. The polarizations of
 H$\beta$ and MgII lines are also shown and labeled as $\beta$ and Mg,
 respectively. The direction of the sky PA 0\degr, 45\degr, 90\degr,
 and 135\degr are also shown. }
 \label{res-4C09.72-qu}
\end{figure}

\begin{figure}
 \includegraphics[width=80mm]{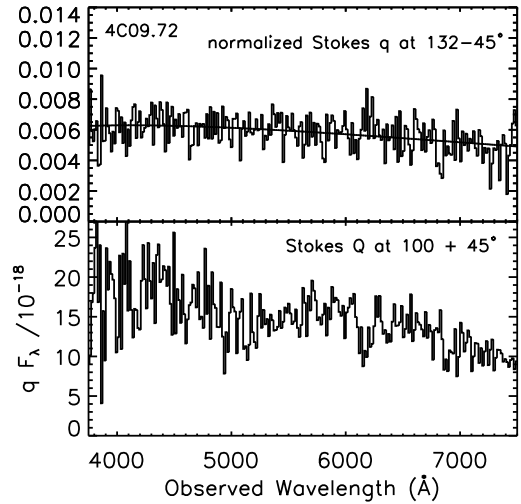}
 \caption{The data for 4C09.72 before any ISP correction. The upper
 panel shows the normalized Stokes parameter $q$ with the reference axis
 at PA = 132\degr\ $-$ 45\degr, i.e. perpendicular (on the $q$-$u$
 plane) to the PA of the quasar's
 intrinsic polarization. The smooth curve is the best fit Serkowski
 curve.
 The lower panel shows the unnormalized Stokes parameter $Q$ with the
 reference axis at PA = 100\degr\ + 45\degr, i.e. perpendicular to the
 PA of the inferred ISP.}
 \label{res-4C09.72-rot-qQ}
\end{figure}

\subsection{4C09.72}

Fig.\ref{res-4C09.72} shows the observed polarization for 4C09.72. The
upper part with four panels shows from top to bottom (1) Stokes $I$
($F_{\lambda}$), (2) normalized $q'$ and $u'$ (thick and thin line,
respectively), (3) unnormalized $Q'$ and $U'$ ($q' \times F_{\lambda}$
and $u' \times F_{\lambda}$) with $I$ (dotted line) scaled to match $Q'$
at the red side, (4) position angle (PA) spectrum. The Stokes parameters
are rotated to have their reference axis along the PA of the
polarization calculated at 4000-4731\AA\ (Table~\ref{tab-pol}; this PA
is indicated by the dotted line in the panel for the PA spectrum in
Fig.\ref{res-4C09.72}), thus designated as $q'$, etc. rather than $q$,
etc. Therefore, $q'$ is essentially the polarization $P$, and $q' \times
F_{\lambda}$ is essentially the polarized flux $P \times
F_{\lambda}$. We summarize the symbols used in this paper in Table
\ref{tab-symbol}.  The separate bottom part shows the polarized flux
spectrum in $\nu F_{\nu}$ form, i.e. $q' \times \nu F_{\nu}$, against
$\lambda$ with both axes in log scale (thus note that x-axes are
different in these two parts). The total flux in $\nu F_{\nu}$ is also
shown (dotted line).  The rest wavelengths are shown at the top of each
part.

\subsubsection{ISP correction~$^1$}

\footnotetext[1]{The reader who wishes to skip to our best-corrected
data can proceed to section \ref{res-4C09.72-bestcor} and
Fig.\ref{res-4C09.72-cor100} (and skip Fig.\ref{res-4C09.72-qu} and
\ref{res-4C09.72-rot-qQ}). Likewise, in the following sections for
each object, the ISP-related discussion is essentially separated in
the first subsection when substantially addressed, so that the reader
who wishes to skip it can proceed directly to the second subsection.}

One distinct observed feature in this object is that the PA rotates
shortward of 4000\AA\ in the rest frame, probably flattens again
shortward of the rest 3500\AA.  We argue that this is most likely due
to a contamination by ISP in our Galaxy and that the data can be
corrected for this contamination to the spectra shown in
Fig.\ref{res-4C09.72-cor100}.

In Fig.\ref{res-4C09.72-qu}, we put the same data shown in
Fig.\ref{res-4C09.72} on the $q$-$u$ plane (normalized Stokes
parameters; binned by 100 pixels or 260\AA). The size of the crosses
shows the error.  The labels are the central rest wavelengths in each
bin in units of 100\AA.  We see that the data points approximately
align linearly, but the line of the alignment does not pass through
the orgin $(q,u)=(0,0)$.  This is what we expect if the observed
polarization consists of an ISP component and an intrinsic
polarization of the object at a fixed PA.  In general, if the
polarization of an object changes in magnitude over the wavelengths
but stays constant in PA, then the data points on the $q$-$u$ plane
align linearly with the alignment direction pointing toward the
origin.  On the other hand, a small contamination from an ISP, to a
first order, will shift all the data points by a constant vector,
since the ISP is in principle fixed at a certain position angle (PA)
on the sky with its magnitude not changing significantly over the
optical wavelengths \citep{SMF75}.  The data points will stay
linearly aligned, but the extrapolation of the alignment direction
will not pass through the origin in this case. This seems to be what
we see in Fig.\ref{res-4C09.72-qu}.

If the PA of the polarization of the emission lines is the same as the
continuum or if the emission lines are intrinsically unpolarized, the
data point for the emission lines should also align along the other
data. This is also the case, as we show in Fig.\ref{res-4C09.72-qu}
the polarization of the H$\beta$ and MgII lines from
Table~\ref{tab-hbeta} and \ref{tab-mg}.

If this two-component interpretation is correct, the alignment
direction of the data points on the $q$-$u$ plane gives us the
intrinsic PA of the polarization of the object.  The ISP can roughly
be inferred from the displacement of the data points: at least the
displacement gives a constraint on the possible PA range of the
ISP. However, the determination of an exact ISP vector requires an
assumption of a PA of the ISP (or a magnitude of the ISP, where two
possible solutions for the PA would exist).  The polarization of
Galactic stars in the same field can be used to check if it is
consistent with the displacement of the data points seen on the
$q$-$u$ plane, and then to assume the PA of the ISP.

The ISP probed by Galactic stars around 4C09.72 is consistent with this
interpretation.  Firstly, our observation of a star which is close to
4C09.72 in projection (see Table \ref{tab-log-star}) shows an ISP of
0.3\% at PA 105\degr. This is overall a reasonable representation for
the displacement of the origin of the quasar's intrinsic polarization at
least in terms of its direction (see Fig.\ref{res-4C09.72-qu}).
Secondly, the observations of several stars around 4C09.72 in the
literature compiled by \citet{He00} show a rather clear tendency for the
PA of the ISP to be at 100\degr - 110\degr with $P = 0.2-0.3$\% (for the
stars of estimated distance of 100-300pc).  Therefore, it is likely that
the observation is simply affected by an ISP and the intrinsic
polarization of 4C09.72 is (at least approximately) at a fixed PA.

For a given assumed PA of the ISP with a Serkowski curve (we also
assume the empirical relation of the constant in the Serkowski-law
with $\lammax$; \citealt{Wh92}), we can determine the best fit
magnitude of the ISP which produces the most flat PA spectrum for the
intrinsic polarization of the object.  Fig.\ref{res-4C09.72-cor100}
shows the result of this correction.  If the PA of the ISP is assumed
to be at 100\degr\ (see below for the case of a larger PA), the best
fit for the 4C09.72 data results in having a maximum polarization of
$\pmax = 0.70 \pm 0.03$\% at $\lammax = 4130 \pm 540$\AA\ for the ISP
and $\thetapa=131.9 \pm 1.7$\degr\ for the intrinsic polarization of
4C09.72 (this PA was already apparent from
Fig.\ref{res-4C09.72-qu}). The Stokes parameters in
Fig.\ref{res-4C09.72-cor100} have been rotated to have their reference
axis to be at this intrinsic PA for the quasar.

We can check if the ISP component in the observed polarization really
resembles a Serkowski curve, by looking at the spectrum of the Stokes
parameter perpendicular in the $q$-$u$ plane (or 45\degr\ in the sky
plane) to the intrinsic polarization of the quasar, since this Stokes
parameter spectrum should not contain the polarization of the
quasar. This is shown in the upper panel of
Fig.\ref{res-4C09.72-rot-qQ}: the spectrum is nicely fit by a Serkowski
curve (this is of course essentially equivalent to checking whether the
ISP-corrected stokes $u$, shown in Fig.\ref{res-4C09.72-cor100}, is
flat).  Similarly, we can check the shape of the ISP-corrected polarized
flux of the object by looking at the Stokes parameter perpendicular in
the $q$-$u$ plane to the PA of the ISP as shown in the lower panel of
Fig.\ref{res-4C09.72-rot-qQ}: this essentially shows the same spectrum
as the $q \flam$ spectra in Fig.\ref{res-4C09.72-cor100}.

We note that the ISP magnitude derived above looks large compared with
those for the nearby star we observed or several other stars in the
literature.  The estimated $\ebv$ for the line of sight to 4C09.72 is
0.042 mag, and thus we might expect that the ISP for 4C09.72 would be
less than $9\ebv \sim 0.4$\% (see Table~\ref{tab-isp}).  However, the
flatness of the PA spectrum after the ISP subtraction
(Fig.\ref{res-4C09.72-cor100}) makes this correction quite believable.
We note again that the limit of $9\ebv$ are less certain for the lines
of sight with smaller $\ebv$ in Figure 9 of \citet{SMF75}.

The EW of the emission lines in the corrected polarized flux is
consistent with zero and its upper limit for the H$\beta$ line is much
less than the EW of the line in the total flux (Tables \ref{tab-hbeta}
and \ref{tab-mg}). This is what we expect since the assumed PA of the
ISP is essentially the same as the observed PA of the line
polarization, and the data points for the lines in the q-u plane are
essentially aligned with other data points. Conversely, our
interpretation is well supported by this line polarization.

The uncertainty in the PA of the ISP leads to an uncertainty in the
corrected polarized flux spectrum.  However, the PA of the ISP is not
likely to be much smaller than 100\degr\ based on the ISP map around
the quasar.  Conversely, if we increase the PA assumed, the features
seen in the corrected polarized flux which are described below become
stronger in this particular case (this makes the magnitude of the
inferred ISP even larger as we can see from Fig.\ref{res-4C09.72-qu}).
Thus, the overall spectral features do not depend on the ISP
correction at least qualitatively.  The uncertainty in the PA of the
ISP may be minimized by the future observations of a few more close-by
distant stars around 4C09.72.

\subsubsection{Corrected spectra}\label{res-4C09.72-bestcor}

Fig.\ref{res-4C09.72-cor100} shows the spectra for 4C09.72 which have
been corrected for the contamination from the ISP in our Galaxy.  Note
that the S/N of the polarization spectra has been effectively degraded
due to the reduction of the $P$ magnitude from the subtraction of the
ISP.

The corrected polarized flux (unnormalized Stokes $Q'$, or $q' \times
\flam$) shows spectral features that look similar to those seen in Ton
202 (Paper I).  Firstly, there are essentially no emission lines,
which are quantified in Table~\ref{tab-hbeta} and
\ref{tab-mg}. Secondly, in the Balmer edge region, there is a slope
down-turn (or local maximum) at about 4000\AA, and there also seems to
be a slope up-turn (or local minimum) at $\sim$ 3600\AA. In addition,
there seems to be a local peak feature at $\sim$3050\AA. These are all
discussed in section \ref{sec-disc}.

\begin{figure}
 \includegraphics[width=80mm]{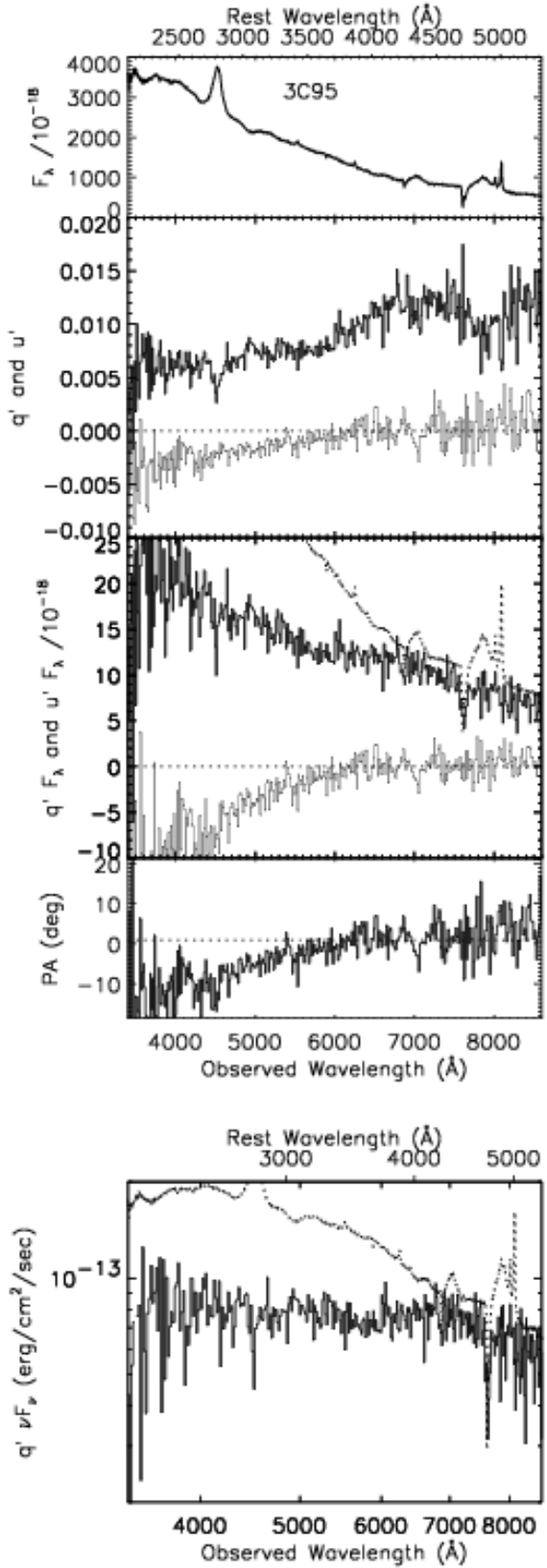}
 \caption{ The same as Fig.\ref{res-4C09.72}, but for 3C95}
 \label{res-3C95}
\end{figure}

\begin{figure}
 \includegraphics[width=80mm]{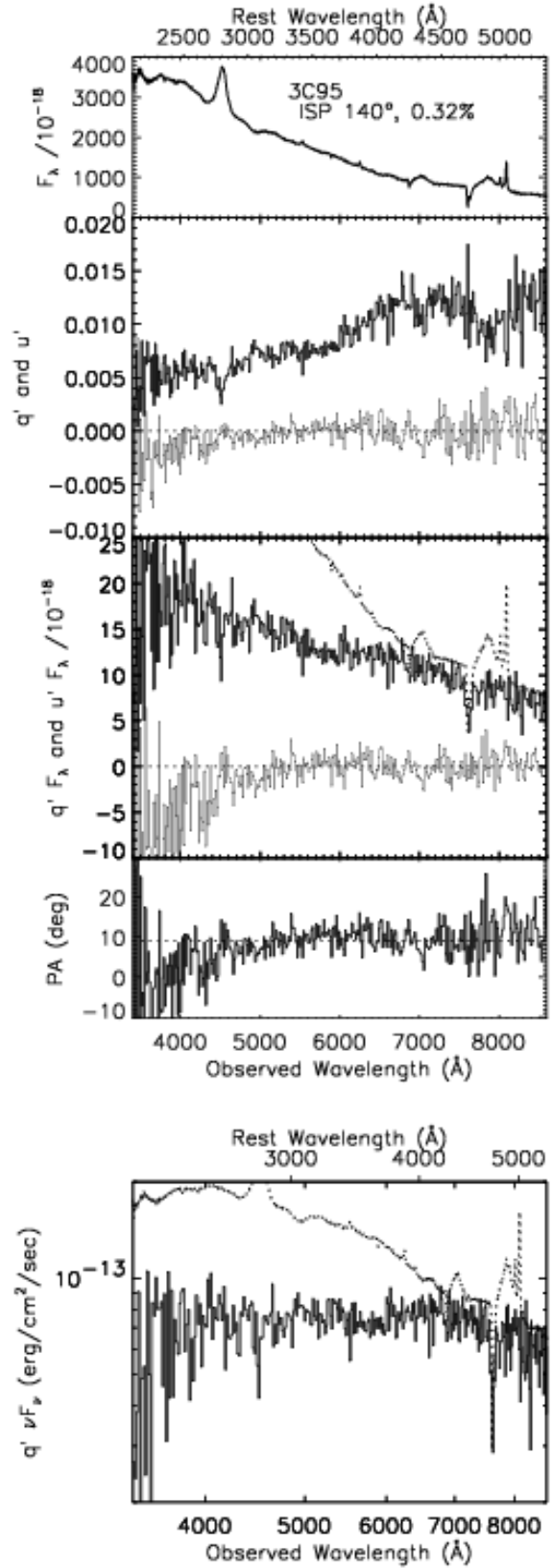}
 \caption{The ISP-corrected polarization of 3C95, assuming ISP PA of 140\degr.}
 \label{res-3C95-cor140}
\end{figure}

\begin{figure}
 \includegraphics[width=80mm]{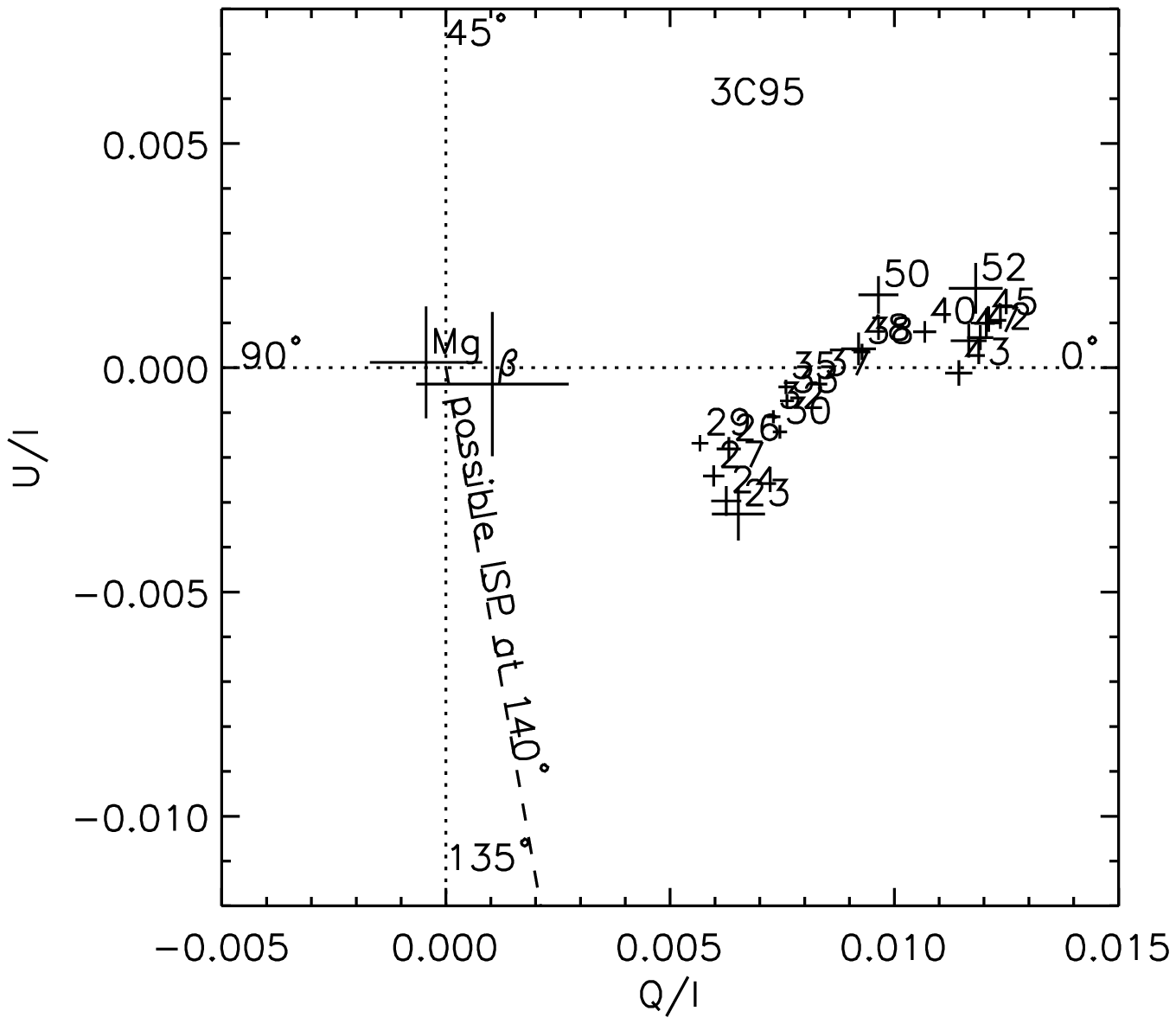}
 \caption{The same as Fig.\ref{res-4C09.72-qu}, but for 3C95.}
 \label{res-3C95-qu}
\end{figure}

\subsection{3C95}

\subsubsection{ISP correction~$^1$}

The observed polarization for 3C95 is shown in Fig.\ref{res-3C95}.  We
have combined the observations with an order sorting filter GG435
(thus $\lambda_{\rm obs} > 4600$\AA) and the ones without any filter
(thus $\lambda_{\rm obs} < 6400$\AA).  There is again a PA rotation
shortward of rest 4000\AA.  This is probably due to a small ISP
contamination and we show the data corrected for the ISP contamination
in Fig.\ref{res-3C95-cor140}. This inference is again based on the
$q$-$u$ plane plot where the data points align at least approximately
linearly as shown in Fig.\ref{res-3C95-qu} (however see below).
The polarization measurements of a few Galactic stars having estimated
distance of 100-600pc \citep{He00} suggests an overall tendency of the
ISP at PA 130-150\degr with $P=0.1-0.4$\%, though our observation of a
neighboring star (Table \ref{tab-log-star}) shows no polarization larger
than 0.1\%.  This direction of the PA seems to be a reasonable
representation for the shift of the intrinsic polarization of 3C95 on
the $q$-$u$ plane (Fig.\ref{res-3C95-qu}).

Thus the argument for ISP contamination looks reasonable.
However, there also seems to be some systematic deviation from the
alignment on the $q$-$u$ plane at the blue end of the spectrum,
shortward of the rest $\sim$3000\AA.  Therefore, we tried to correct
for the ISP using the data longward of this wavelength.  The same
correction procedure as for 4C09.72, assuming the PA of the ISP as
140\degr, produced the results shown in Fig.\ref{res-3C95-cor140}.
The fit resulted in having $\pmax = 0.32 \pm 0.03$\% and the PA of the
quasar's intrinsic polarization at $9.7 \pm 0.9$\degr\ ($\lammax$ was
simply assumed to be 5500\AA).  The correction looks overall
reasonable except for the blue end (the ISP correction using the whole
wavelength results in having some wave-like rotation in the PA
spectrum).

There is an uncertainty in the polarized flux shape due to the
uncertain ISP PA, but this is rather small since the magnitude of the
ISP is small compared to the observed polarization.  If the PA of the
ISP is assumed to be smaller than 140\degr, this would give a larger
ISP magnitude as can be seen in Fig.\ref{res-3C95-qu}, which is
unlikely as we argued above.  For an ISP correction with a
larger PA, the spectral features in the polarized flux even become
slightly stronger.

The H$\beta$ and MgII lines in the spectrum before any ISP correction
are unpolarized with $\sigma \sim 0.15\%$ (see Table \ref{tab-hbeta}
and \ref{tab-mg}).  This means that the ISP contamination
is small (0.3-0.4\% at most) and the lines are intrinsically
unpolarized, unless the ISP is coincidentally canceling out any
intrinsic polarization component. (The above ISP correction 'produces'
a little line polarization, but not significantly).

\subsubsection{Corrected spectra}

Fig.\ref{res-3C95-cor140} shows the data corrected for the ISP
contamination.  The polarized flux spectrum again shows a spectral
feature at the Balmer edge region similar to that of Ton 202 in Paper I
and 4C09.72 in the previous section: a slope down-turn at around rest
4000\AA\ and an up-turn at around 3600\AA.  As we argued above, the
emission lines probably are intrinsically unpolarized, even though the
[OIII] line and MgII lines look marginally polarized in the corrected
polarized flux spectrum.

\begin{figure}
 \includegraphics[width=80mm]{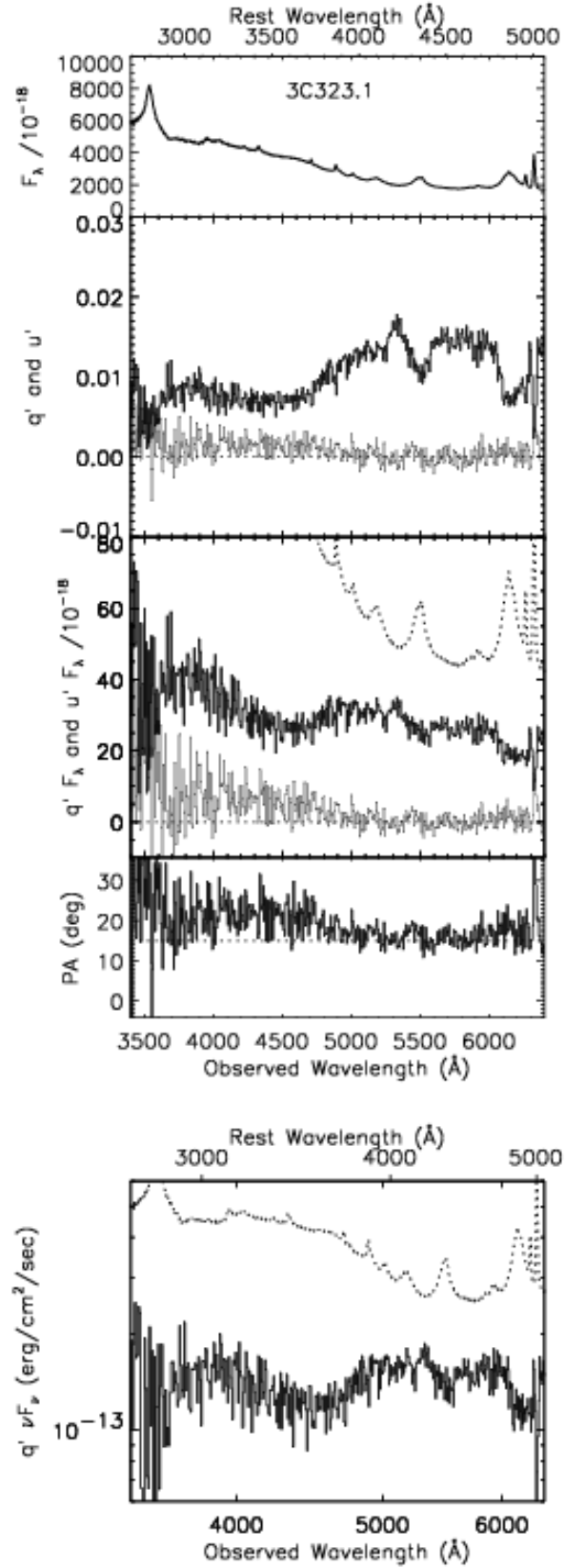}
 \caption{ The same as Fig.\ref{res-4C09.72}, but for the Keck
 observation of 3C323.1.}
 \label{res-3C323.1}
\end{figure}

\begin{figure}
 \includegraphics[width=80mm]{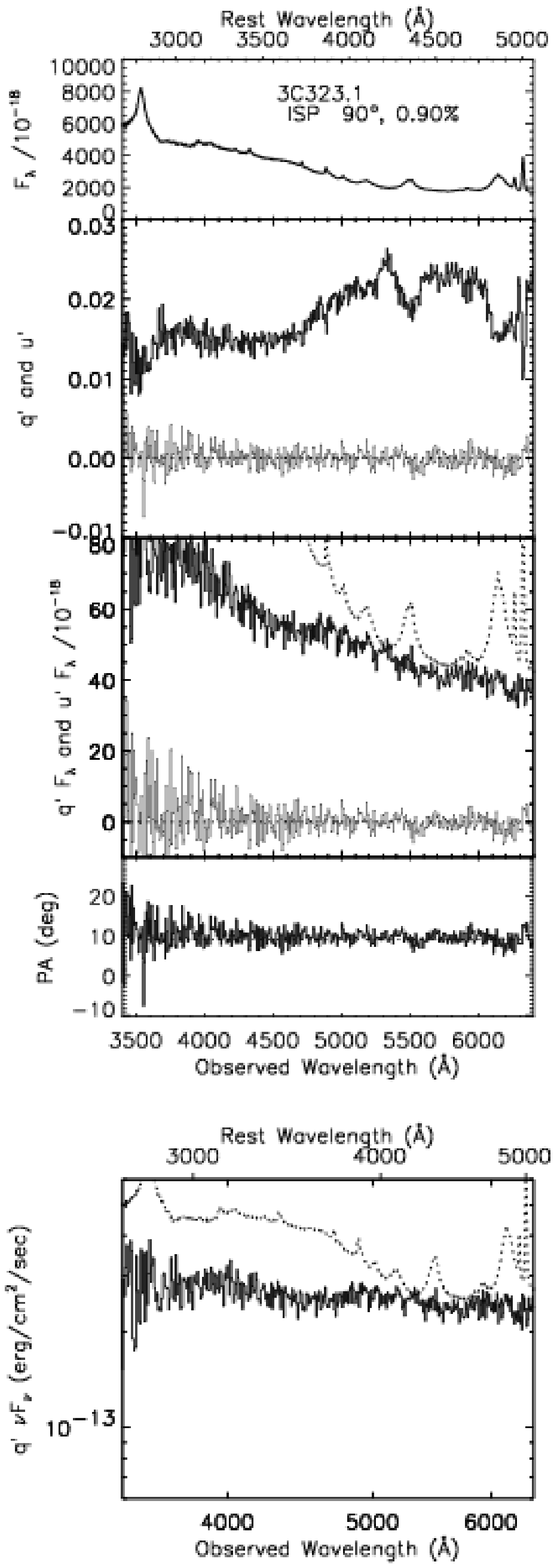}
 \caption{The ISP-corrected polarization of 3C323.1, assuming ISP PA of 90\degr.}
 \label{res-3C323.1-cor090}
\end{figure}

\begin{figure}
 \includegraphics[width=80mm]{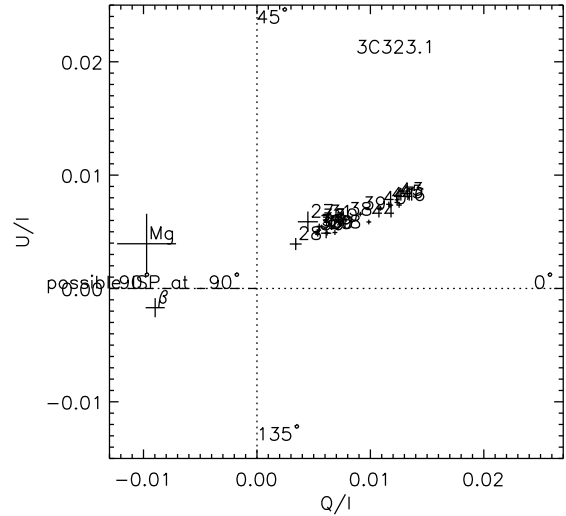}
 \caption{The same as Fig.\ref{res-4C09.72-qu}, but for 3C323.1.}
 \label{res-3C323.1-qu}
\end{figure}

\subsection{3C323.1}

\subsubsection{ISP correction~$^1$}

The data for this object, shown in Fig.\ref{res-3C323.1}, are also
likely to have been affected by ISP: the data corrected for the ISP
are shown in Fig.\ref{res-3C323.1-cor090}.

The original measurement in Fig.\ref{res-3C323.1} shows a PA rotation
shortward of rest 4000\AA\ and also at emission line wavelengths, but
on the $q$-$u$ plane, a linear alignment of the data points is seen
(Fig.\ref{res-3C323.1-qu}).  The direction of the shift of the data
points is consistent with the observations of Galactic stars in the
same field.  Our observation of a nearby star shows $P=0.7$\% at PA
84\degr\ (see Table \ref{tab-log-star}), and other observations of
three Galactic stars within 3\degr\ \citep{He00} having estimated
distance larger than 200pc also show a similar PA (85-95\degr), though
with a smaller P (0.2-0.3\%).  Therefore we have corrected the data
assuming the PA of the ISP at 90\degr, and
Fig.\ref{res-3C323.1-cor090} is the result of this correction.  Before
this ISP correction, there were some apparent features at the emission
line wavelengths in the polarized flux (Fig.\ref{res-3C323.1}), but
after the correction, these are essentially gone.  This is expected
since the polarization measured for these lines before the correction
is roughly consistent with the subtracted ISP
(Fig.\ref{res-3C323.1-qu}; Tables \ref{tab-hbeta}, \ref{tab-mg}).

The uncertainty in the PA of the ISP leads to the uncertainty in the
polarized flux.  An ISP correction with a smaller PA results in a
stronger spectral feature in the polarized flux. With a larger PA, the
feature becomes weaker, but this also results in having emission lines
in the polarized flux.  The ISP magnitude also increases rapidly with
a larger assumed PA.

\subsubsection{Corrected spectra}

The data which are best-corrected for the ISP are shown in
Fig.\ref{res-3C323.1-cor090}. The emission lines look unpolarized (see
also Table \ref{tab-hbeta} and \ref{tab-mg}) after the ISP correction.
At the Balmer edge region, there remains the now-familiar
intrinsic features left: the slope down-turn at $\sim
4000$\AA, and an up-turn at $\sim 3600$\AA, though the feature looks
weaker than for the previous objects.

\begin{figure}
 \includegraphics[width=80mm]{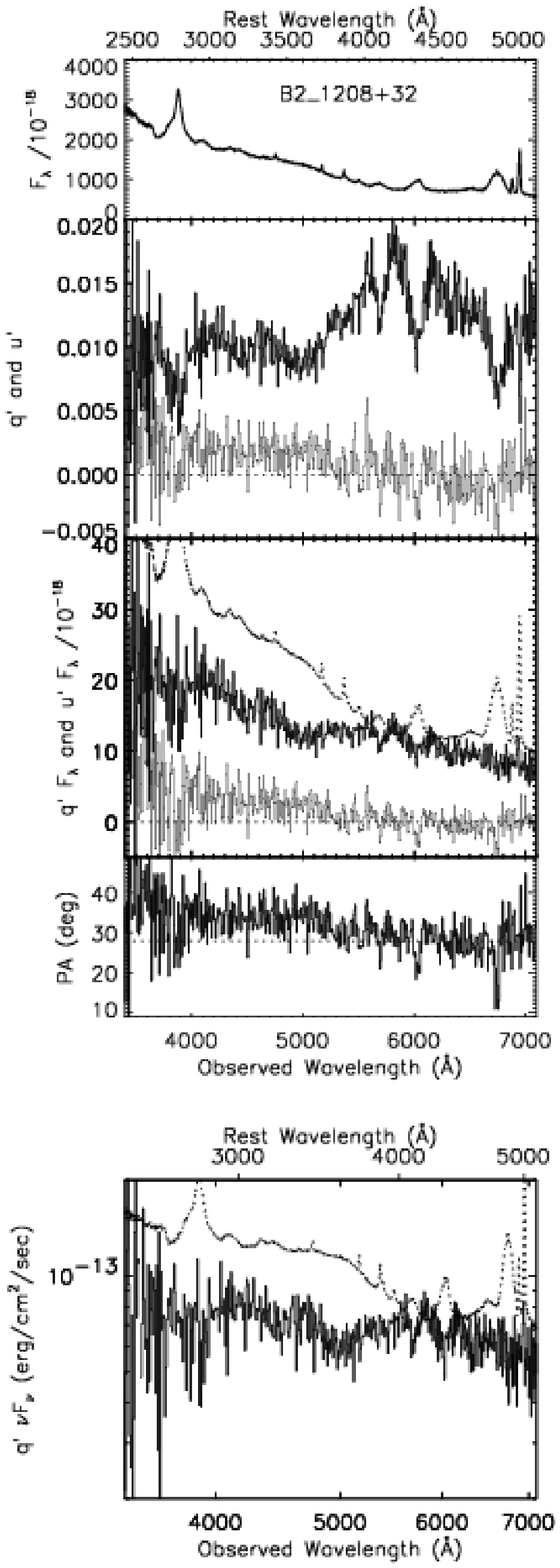}
 \caption{ The same as Fig.\ref{res-4C09.72}, but for the Keck
 observation of B2 1208+32.}
 \label{res-B2_1208+32}
\end{figure}

\begin{figure}
 \includegraphics[width=80mm]{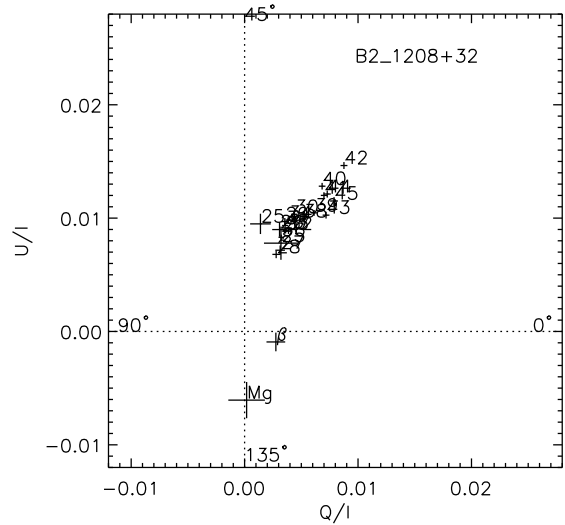}
 \caption{The same as Fig.\ref{res-4C09.72-qu}, but for B2 1208+32.}
 \label{res-B2_1208+32-qu}
\end{figure}

\subsection{B2 1208+32}

\subsubsection{ISP effect~$^1$}

In this object, a small PA rotation shortward of 4000\AA\ is again seen
(Fig.\ref{res-B2_1208+32}) and the data points on the $q$-$u$ plane also
roughly align linearly (Fig.\ref{res-B2_1208+32-qu}), though not too
cleanly. We do not have our own observation of a Galactic star in the
field, but the data in the literature indicate that stars with
estimated distances of 400-500pc show a small polarization ($\la 0.25$\%)
at PA 0-25\degr.

If there is any effect of ISP around at this PA, it is expected to
be small, judging from the location of the data points on the $q$-$u$
plane.  However, we note that the continuum PA rotation direction and
the line polarization direction are not in the same sense,
which suggests that one of them is intrinsic.  In
either case, the ISP magnitude is probably small, so its effect should
also be small.

\subsubsection{Uncorrected spectra}     

Fig.\ref{res-B2_1208+32} shows the measurements without any ISP
correction, and the intrinsic polarization is expected to be
qualitatively the same, since the ISP effect is thought to be small.
The measured small line polarization (polarized at different PA from
the continuum) could either be intrinsic or entirely due to an ISP.
The polarized flux shows a similar, now-familiar feature in the Balmer
edge region, and the up-turn at $\sim 3600$\AA\ in this object looks
more pronounced.

\begin{figure}
 \includegraphics[width=80mm]{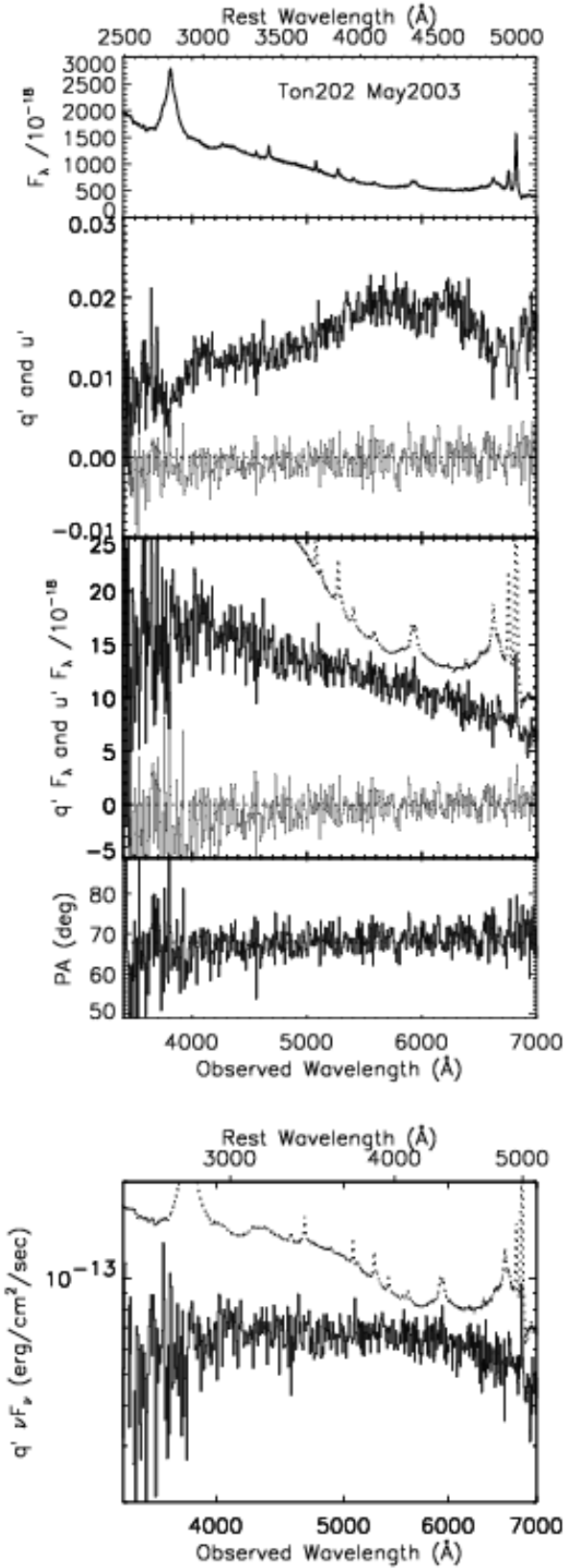}
 \caption{ The same as Fig.\ref{res-4C09.72}, but for the Keck
 observation of Ton202.}
 \label{res-Ton202}
\end{figure}

\begin{figure}
 \includegraphics[width=80mm]{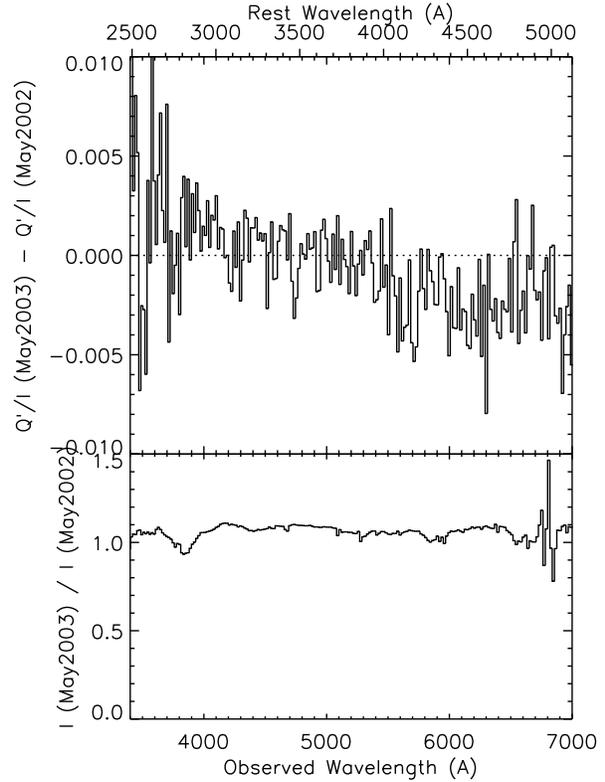}
 \caption{Comparison of the observations of Ton202 at two epochs. The
 upper panel show the polarization observed in May 2003 minus that in
 May 2002. The lower panel show the total flux in May 2003 divided by
 that in May 2002.}
 \label{Ton202-dif}
\end{figure}

\subsection{Ton202}

This object was found to show a Balmer edge feature in the polarized
flux, when observed in May 2002 (Paper I). However, we do not find
such a dramatic feature in our data taken in May 2003, as we show 
in Fig.\ref{res-Ton202}. The polarized flux looks quite
featureless at $\sim 4000$\AA. (At the red end wavelengths longer than
$\sim$6400\AA, the polarized flux shape become convex, which could be
due to second-order light.) This difference in two
epochs seems to be due to a real variation in polarization.

The upper panel of Fig.\ref{Ton202-dif} shows the difference in
polarization in these two epochs (more accurately, difference in
normalized Stokes $q$ rotated to the object's polarization PA at each
epoch; note that the sky PA of the May 2002 observation could not be
determined due to some calibration problem; see Paper I).  Thus the
polarization seems to have decreased longward of $\sim 3700$\AA, while
the continuum flux shape stayed roughly the same as shown in the lower
panel of Fig.\ref{Ton202-dif} (absolute flux levels have uncertainties
from slit loss and seeing size difference, which are rather hard to
quantify).

Polarization variability on a timescale of one year is not
unexpected, since the polarization is thought to originate from a very
small spatial scale, smaller than the size scale of the
broad line region, as we argue in Paper I and also in the next
section. In fact, polarization variability of Ton 202 and some other
quasars in our sample has been pointed out by \citet{SS00}.  Some
broad-line radio galaxies are also known to show polarization
variability \citep{An84}.  Ton 202, as well as other objects, should be
followed up to confirm and explore the variability.

We note that the [OIII]$\lambda5007$ line might be weakly polarized. The
measurements in two epochs gave consistent results for the magnitude of
polarization (Table~\ref{tab-hbeta}). Assuming that this [OIII] line
polarization did not vary, we would be able to constrain that the PA of
the continuum polarization has changed only less than $\sim 9$\degr
(1$\sigma$) based on the 1$\sigma$ uncertainty for the [OIII] line
polarization PA (Table \ref{tab-hbeta}).  There might be a very small
rotation in PA across 4000\AA\ (Fig.\ref{res-Ton202}), which was not
seen in the data of May 2002, but no useful ISP measurement is available
at this time.

\begin{figure}
 \includegraphics[width=80mm]{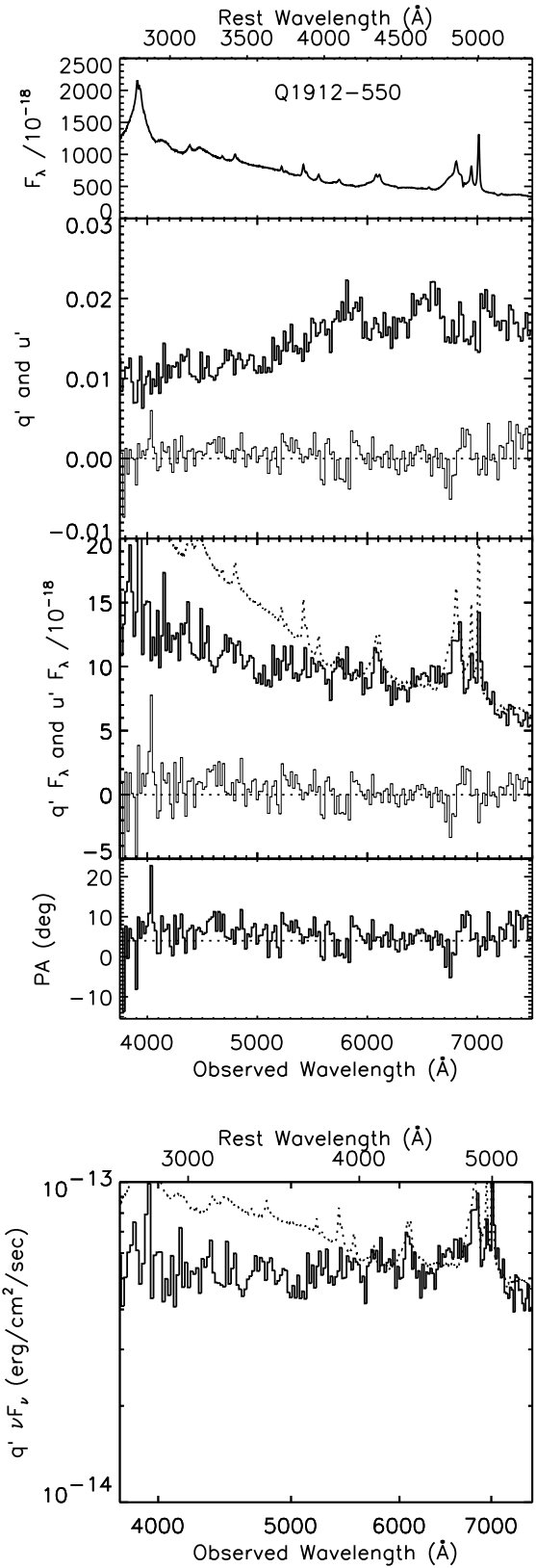}
 \caption{ The same as Fig.\ref{res-4C09.72}, but for Q1912-550.}
 \label{res-Q1912-550}
\end{figure}

\begin{figure}
 \includegraphics[width=80mm]{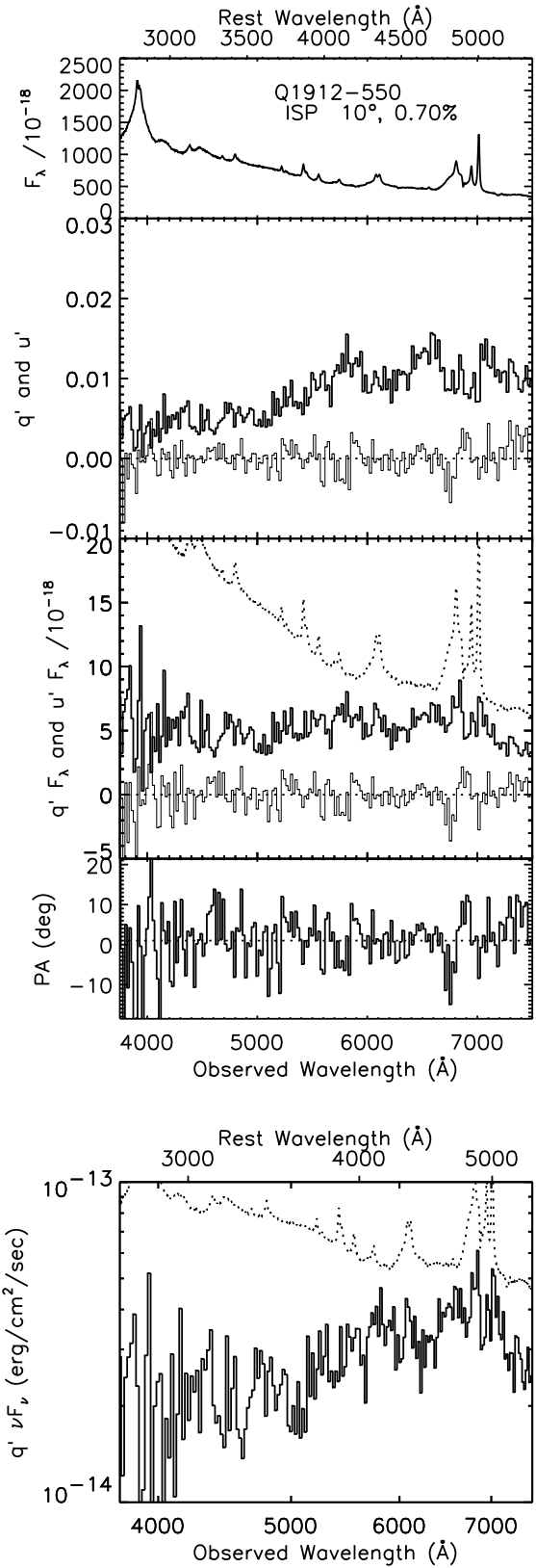}
 \caption{The ISP-corrected polarization of Q1912-550.}
 \label{res-Q1912-550-cor010}
\end{figure}

\begin{figure}
 \includegraphics[width=80mm]{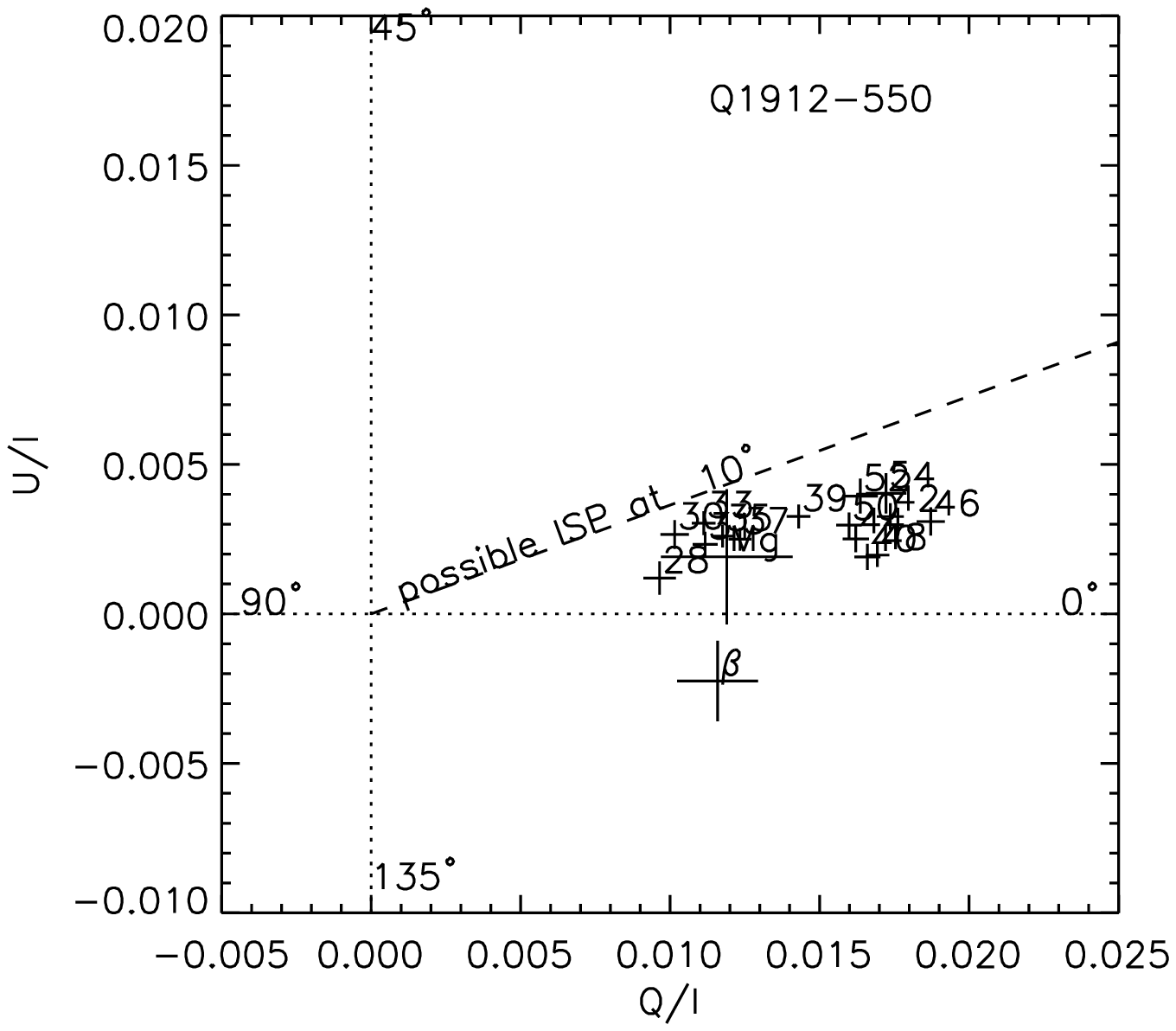}
 \caption{The same as Fig.\ref{res-4C09.72-qu}, but for Q1912-550.}
 \label{res-Q1912-550-qu}
\end{figure}

\begin{figure}
 \includegraphics[width=80mm]{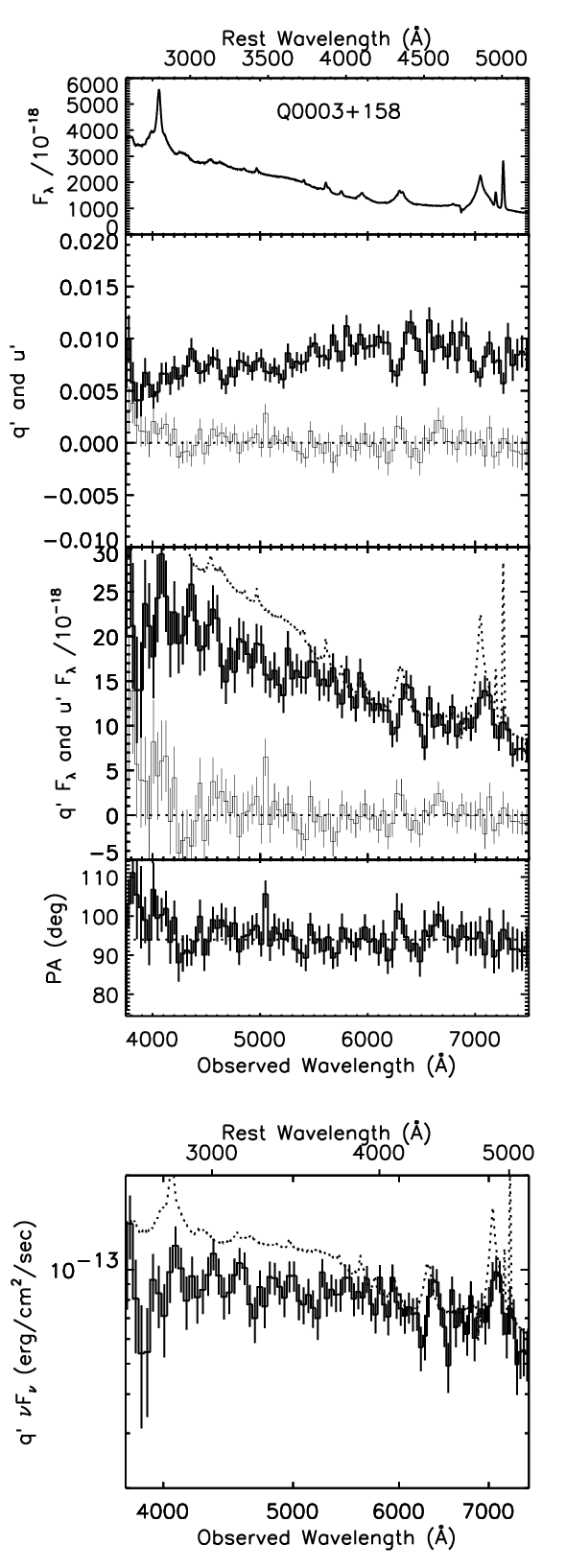}
 \caption{ The same as Fig.\ref{res-4C09.72}, but for Q0003+158.}
 \label{res-Q0003+158}
\end{figure}

\begin{figure}
 \includegraphics[width=80mm]{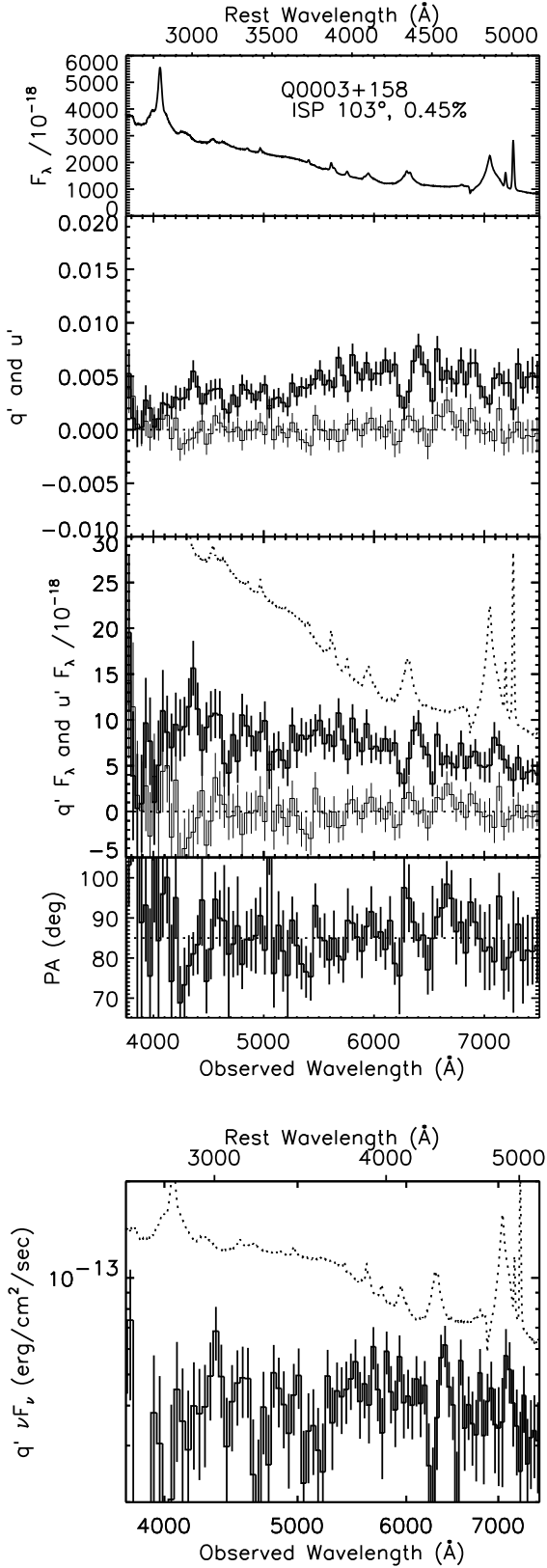}
 \caption{The ISP-corrected polarization of Q0003+158.}
 \label{res-Q0003+158-cor103}
\end{figure}

\begin{figure}
 \includegraphics[width=80mm]{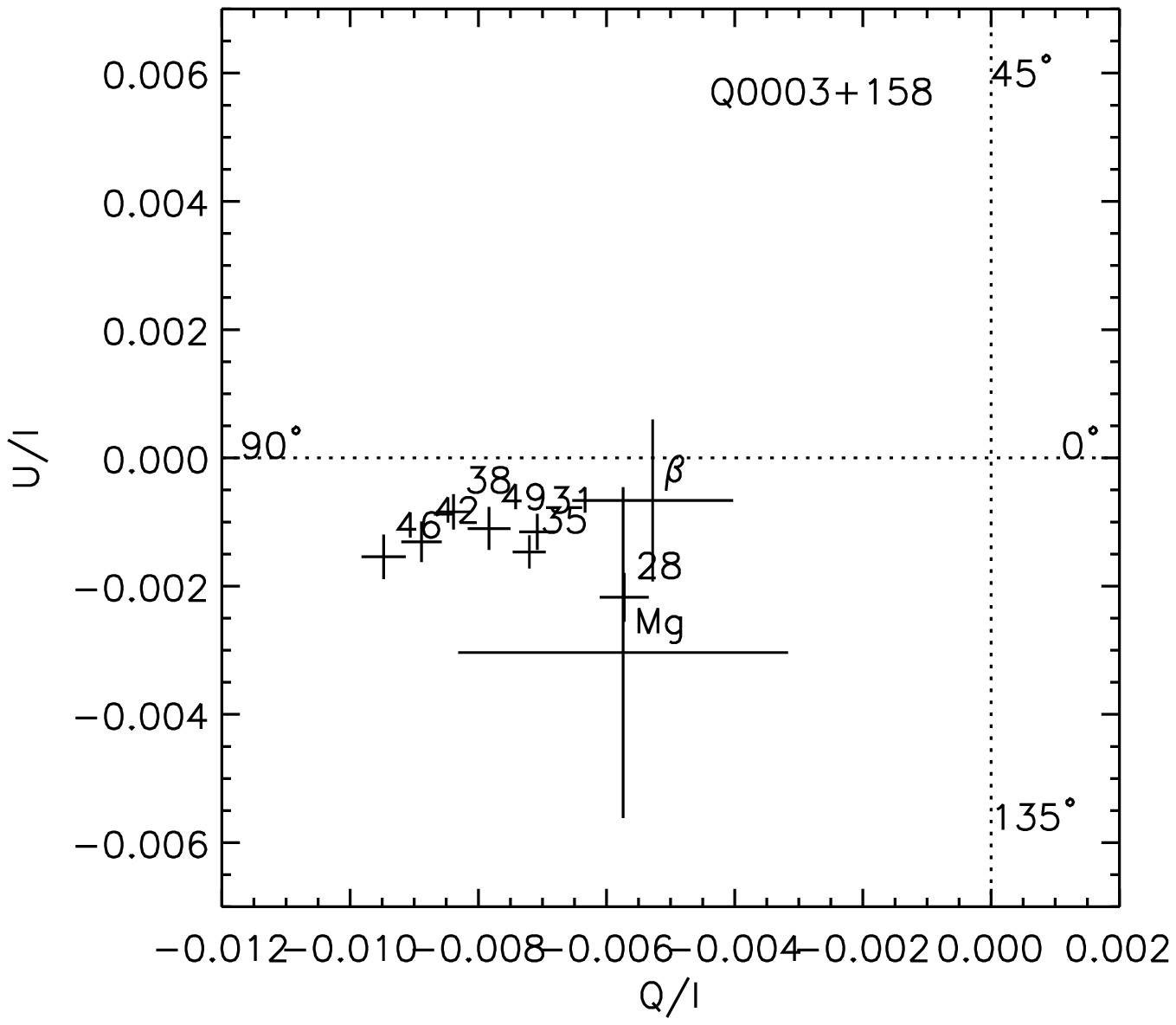}
 \caption{The same as Fig.\ref{res-4C09.72-qu}, but for Q0003+158.}
 \label{res-Q0003+158-qu}
\end{figure}

\begin{figure}
 \includegraphics[width=80mm]{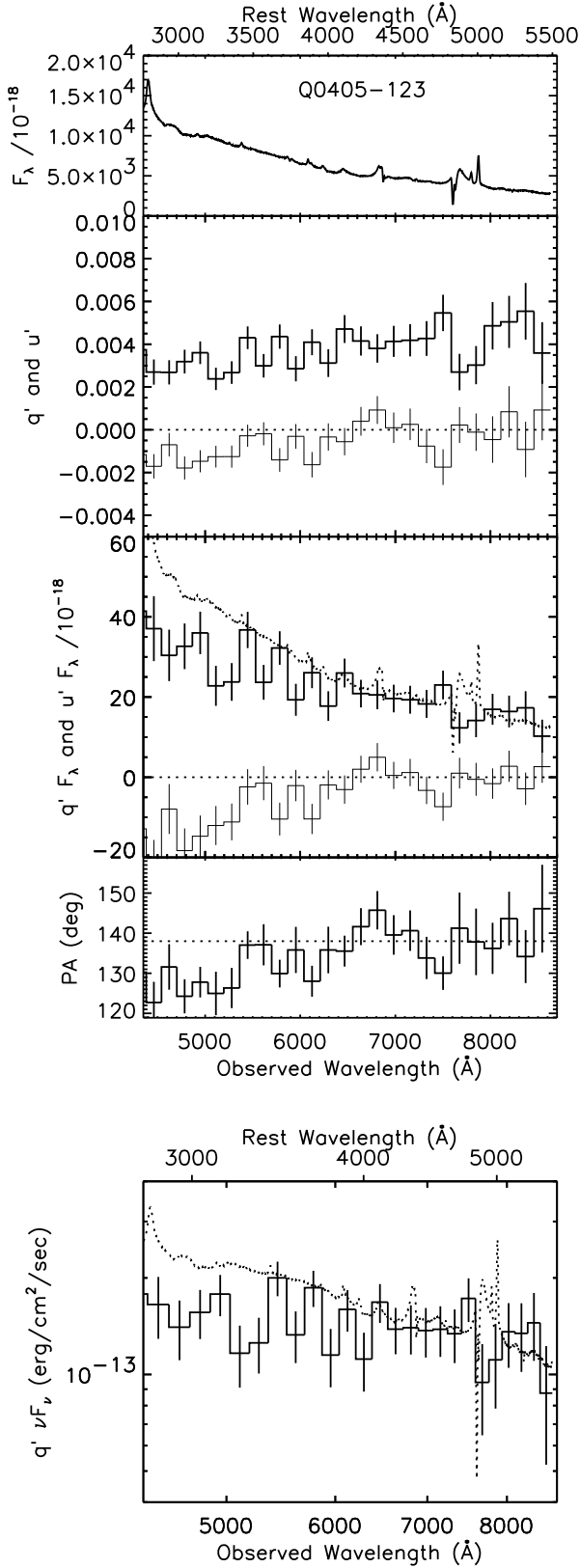}
 \caption{ The same as Fig.\ref{res-4C09.72}, but for Q0405-123.}
 \label{res-Q0405-123}
\end{figure}

\begin{figure}
 \includegraphics[width=80mm]{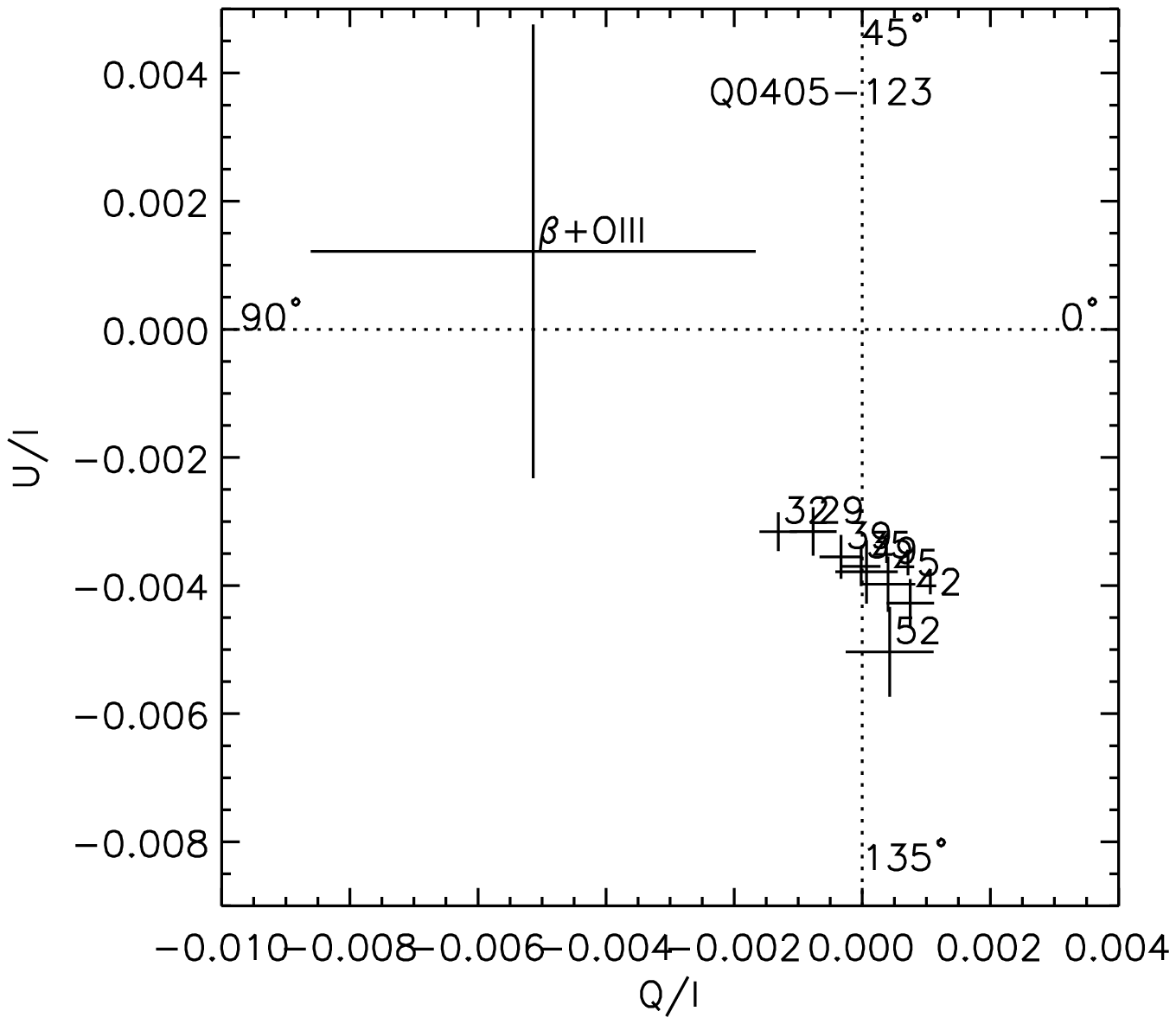}
 \caption{The same as Fig.\ref{res-4C09.72-qu}, but for Q0405-123.  We
 calculated and plotted the line polarization of H$\beta$+[OIII] region
 (4731-5032\AA) instead of H$\beta$ only, since excluding the telluric
 absorption wavelengths at around 7600\AA\ leaves a very small wavelength
 window for H$\beta$ line.}
 \label{res-Q0405-123-qu}
\end{figure}

\subsection{Q1912-550}

\subsubsection{ISP correction~$^1$}

As shown in Fig.\ref{res-Q1912-550}, the polarization clearly
decreases shortward of rest 4000\AA, so the small blue bump is at
least partially unpolarized. However, the H$\beta$ and [OIII] lines
are polarized as seen in the $q' \times \flam$ spectrum.  The data are
probably affected by the ISP, and we show the ISP-corrected data in
Fig.\ref{res-Q1912-550-cor010}.  The data points in the $q$-$u$ plane
shown in Fig.\ref{res-Q1912-550-qu} align roughly linearly, suggesting
the PA of the intrinsic polarization is roughly 0-10\degr.  Our
observation of a nearby star shows a rather strong ISP of $\sim$0.7\%
at PA $\sim$2\degr\ (Table \ref{tab-log-star}).  Therefore, the
observation of the quasar is probably affected by the ISP which
happens to be at a PA similar to that of the object polarization,
causing very little PA rotation in the observed data.

The Galactic latitude is not high for this object, though the          
estimated $\ebv$ is not much larger than other objects (see                  
Table~\ref{tab-isp}).  The existing polarization measurements of the
nearby Galactic stars (within $\sim$ 6\degr; \citealt{He00})
show that, while two stars have low $P$ ($\la 0.15$\%) at PA 70-80\degr
(with the estimated distance of 90-250pc), three stars have a rather
high $P$ of 0.4-0.6\% at PA 5-15\degr (distance 170-440pc). This is
consistent with our observation of one Galactic star.

Fig.\ref{res-Q1912-550-cor010} is the result after the subtraction of
ISP assuming PA=10\degr (with the same $\lammax$ as the observed
star in Table~\ref{tab-log-star}).  In this case, the best fit ISP
magnitude was of $\pmax = 0.7$\%.  With this correction, the narrow
lines are rather well removed from the polarized flux spectrum, but
there appear to be broad components (these were seen before the
correction), which is quite different from the cases of 4C09.72 and
3C95 or 3C323.1.  The ISP magnitude could be slightly larger, if its
PA is smaller than 10\degr (though the magnitude would not be
constrained well when it is too close to the PA of the intrinsic
quasar polarization);  the broad lines still remain even with this
larger ISP subtraction.  In fact, this seems to be due to a
qualitative difference in the original polarization spectrum, rather
than due to undersubtraction of the ISP --- unlike those seen in
4C09.72 (and in 3C95, 3C323.1, and B2 1208+32), the polarization does not
seem to decrease at the wavelengths of broad line wings. Therefore,
the broad lines seem to be slightly polarized.

Since the ISP PA seems to be very close to the PA of the quasar
polarization, this leaves a corresponding rather large uncertainty in
the broad-band shape of the polarized flux.

\subsubsection{Corrected spectra}

Fig.\ref{res-Q1912-550-cor010} shows the ISP-corrected spectra. In
this object, the broad line components seem to be slightly polarized.
One would then expect the small blue bump also to be somewhat
polarized. Therefore, the decline of the intrinsic polarized flux
continuum shortward of 4000\AA\ could be sharper than seen in the
corrected polarized flux.  However, we note that the uncertainty in
the broad-band shape of the corrected polarized flux for this object
is rather large.

\subsection{Q0003+158}

\subsubsection{ISP correction~$^1$}

This object (Fig.\ref{res-Q0003+158}) also has an at least partially
unpolarized small blue bump, since $P$ decreases shortward of 4000\AA,
but the broad lines look partially polarized.  There is no clear evidence
of PA rotations.  Based on the polarization measurements of two
Galactic stars around this object (0.06/0.41\% for 140/280pc at
115/118\degr; \citealt{He00}), there could be a small ISP at a PA very
similar to that observed for Q0003+158.  Therefore, at least
partially, the polarized lines could be due to ISP.  In
Fig.\ref{res-Q0003+158-cor103}, we illustrate a possible ``nominally''
maximum effect of an ISP correction: namely we implemented the same
fitting procedure assuming $\pmax=0.45$\% which is equal to $9\ebv$,
instead of assuming the PA of the ISP.  In these corrected data, the
polarized flux spectrum shows almost no emission lines (i.e. the
emission line polarization seen in the uncorrected data is consistent
with the assumed ISP; see Table~\ref{tab-hbeta} and \ref{tab-mg} as
well as the data on the $q$-$u$ plane shown in
Fig.\ref{res-Q0003+158-qu}).

\subsubsection{Corrected Spectra}

Fig.\ref{res-Q0003+158-cor103} shows the ISP-corrected spectra, but
note that we do not have a firm basis for the ISP estimate for this
object. In these corrected spectra, almost no emission lines are seen
in the polarized flux, and at the Balmer edge region, there seems to
be again a slope down-turn at rest $\sim$4000\AA.  However, we need
more data to confirm these, and also need more data on
Galactic stars around this object.

\subsection{Q0405-123}

\subsubsection{ISP effect~$^1$}

A PA rotation across 4000\AA\ is seen (Fig.\ref{res-Q0405-123}) which
resembles that of 4C09.72 (and 3C95 or 3C323.1).  This could be due to
ISP since the data points in the $q$-$u$ plane roughly align
linearly, with the alignment line displaced from the origin
(Fig.\ref{res-Q0405-123-qu}), although no good polarization
measurements for Galactic stars in the quasar field are available.  The
alignment direction suggests the intrinsic polarization is at PA $\sim
160$\degr.  The line polarization is at least consistent with this
interpretation, but does not provide significant limits
(Fig.\ref{res-Q0405-123-qu}; the polarization of the H$\beta$ +
[OIII]$\lambda$5007 line measured at 4731-5032\AA\ is plotted instead
of H$\beta$ only, since excluding the telluric absorption wavelengths
at around 7600\AA\ left a very small wavelength window for H$\beta$).

\subsubsection{Uncorrected spectra}

The data without any ISP correction are shown in
Fig.\ref{res-Q0405-123}.  The overall polarization is very low
(Fig.\ref{res-Q0405-123}; note the small $P$ of $\sim 0.5$\% even at
the red side), but we marginally detected a polarization decrease
shortward of 4000\AA\ (Table~\ref{tab-pol}).  The above consideration
of a possible ISP effect might suggest that the intrinsic
polarization of this object is at PA $\sim 160$\degr. The lines are
consistent with being unpolarized, but the constraints are not tight.

The object shows variability, as discussed by \citet{SS00}: when
observed in 1990-1991 by \citet{An96}, the optical polarization was much
lower than we observed ($0.04 \pm 0.02$\%). The measurement by
\citet{SS00} in 1995 and 1999 did not show significant optical
polarization ($P=0.27\pm0.19$\% and $0.17\pm0.14$\%, respectively,
though still consistent with our measurement. The measurements by
\cite{SMA84} in 1978 and 1980 are roughly consistent with ours, or $P$
was slightly higher in one occasion in 1978.

\begin{figure}
 \includegraphics[width=80mm]{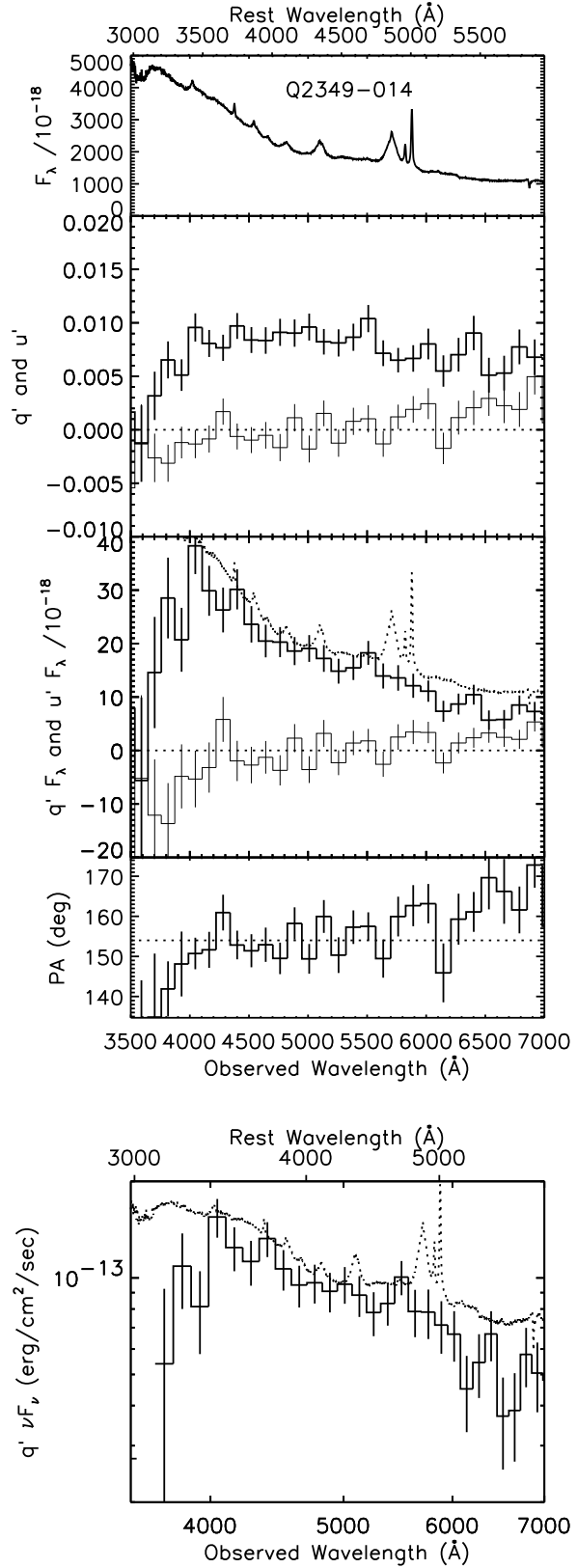}
 \caption{ The same as Fig.\ref{res-4C09.72}, but for Q2349-014.}
 \label{res-Q2349-014}
\end{figure}

\subsection{Q2349-014} 

Although the formal polarization ratio $r_P$ is less than unity at 3
$\sigma$ (Table~\ref{tab-pol}), the polarization seems to behave
differently from the other objects (Fig.\ref{res-Q2349-014}) in that it
looks constant over 4000\AA, but rather abruptly decreases shortward of
$\sim 3500$\AA.  This interesting behavior should be confirmed and
followed up with longer integrations.  There could be a small effect
from an ISP, since the polarization measurements of several Galactic
stars show a rather clear PA alignment at 120-130\degr\ with
$P=0.2-0.4$\%.

\begin{figure}
 \includegraphics[width=80mm]{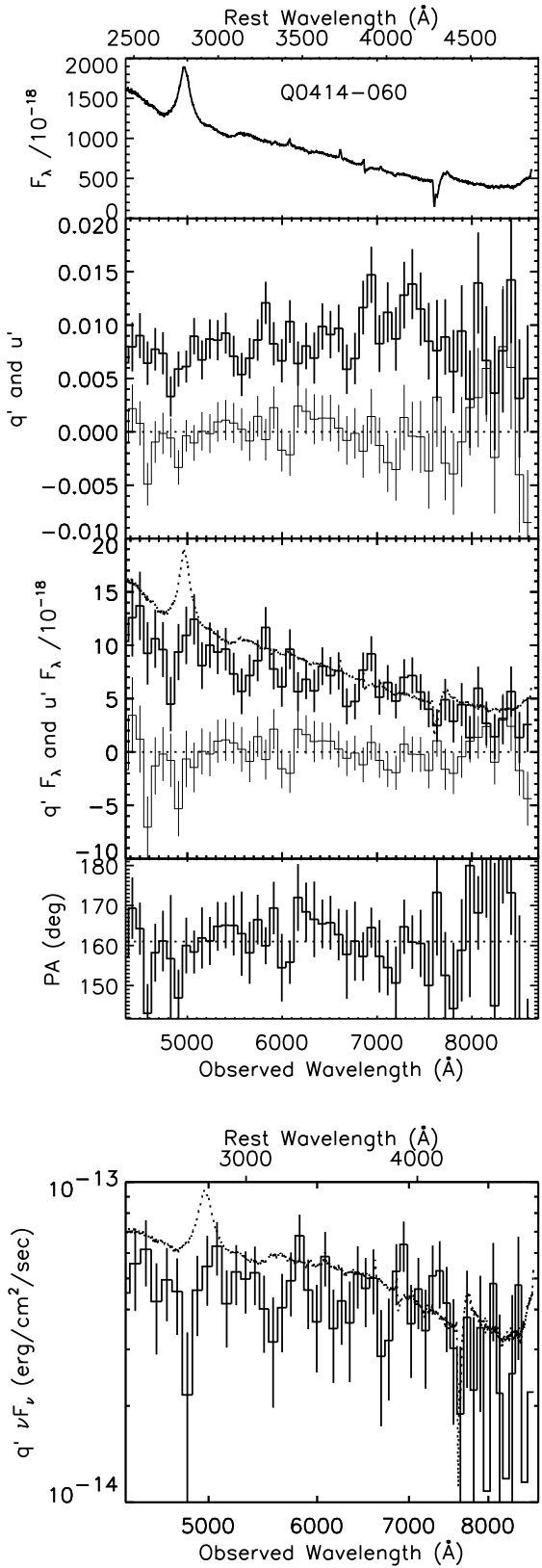}
 \caption{ The same as Fig.\ref{res-4C09.72}, but for Q0414-060.}
 \label{res-Q0414-060}
\end{figure}

\subsection{Q0414-060} 

We did not detect a significant $P$ decrease shortward of 4000\AA\
(Fig.\ref{res-Q0414-060}), and the limit on the MgII line
polarization is not significant either (Table~\ref{tab-mg}).  No good
polarization measurement for Galactic stars is found in the literature
within 5\degr\ from the quasar.

\begin{figure}
 \includegraphics[width=80mm]{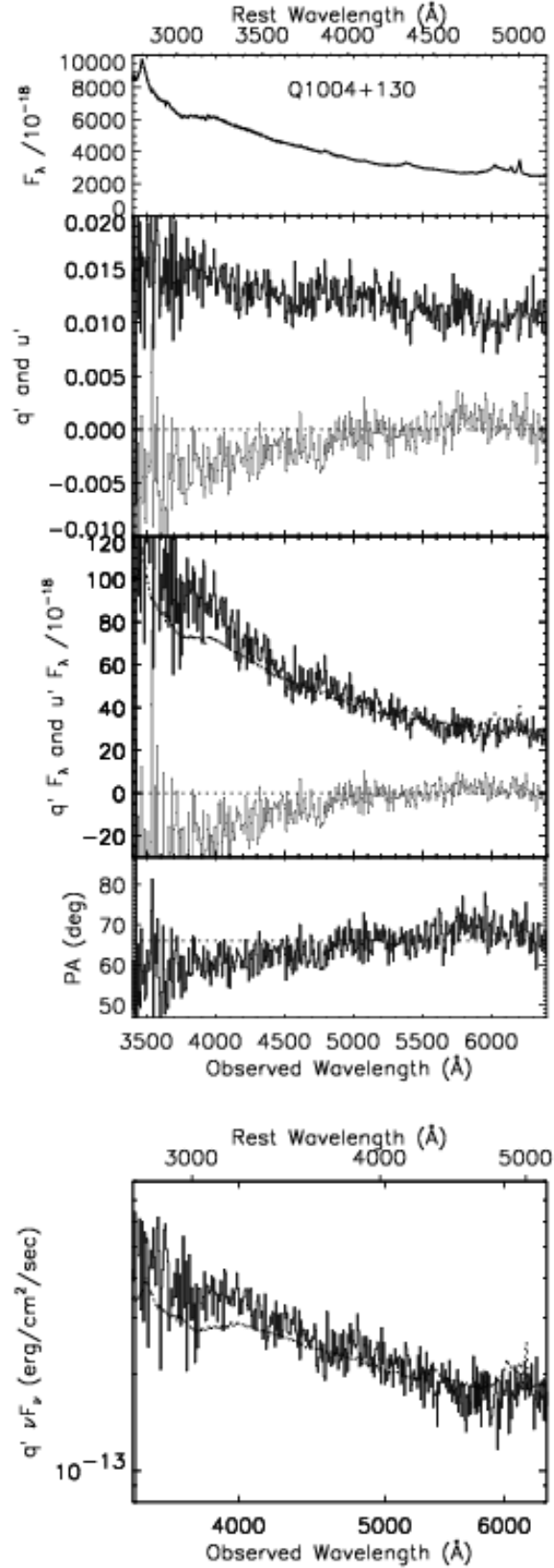}
 \caption{ The same as Fig.\ref{res-4C09.72}, but for the Keck
 observation of Q1004+130.}
 \label{res-Q1004+130}
\end{figure}

\begin{figure}
 \includegraphics[width=80mm]{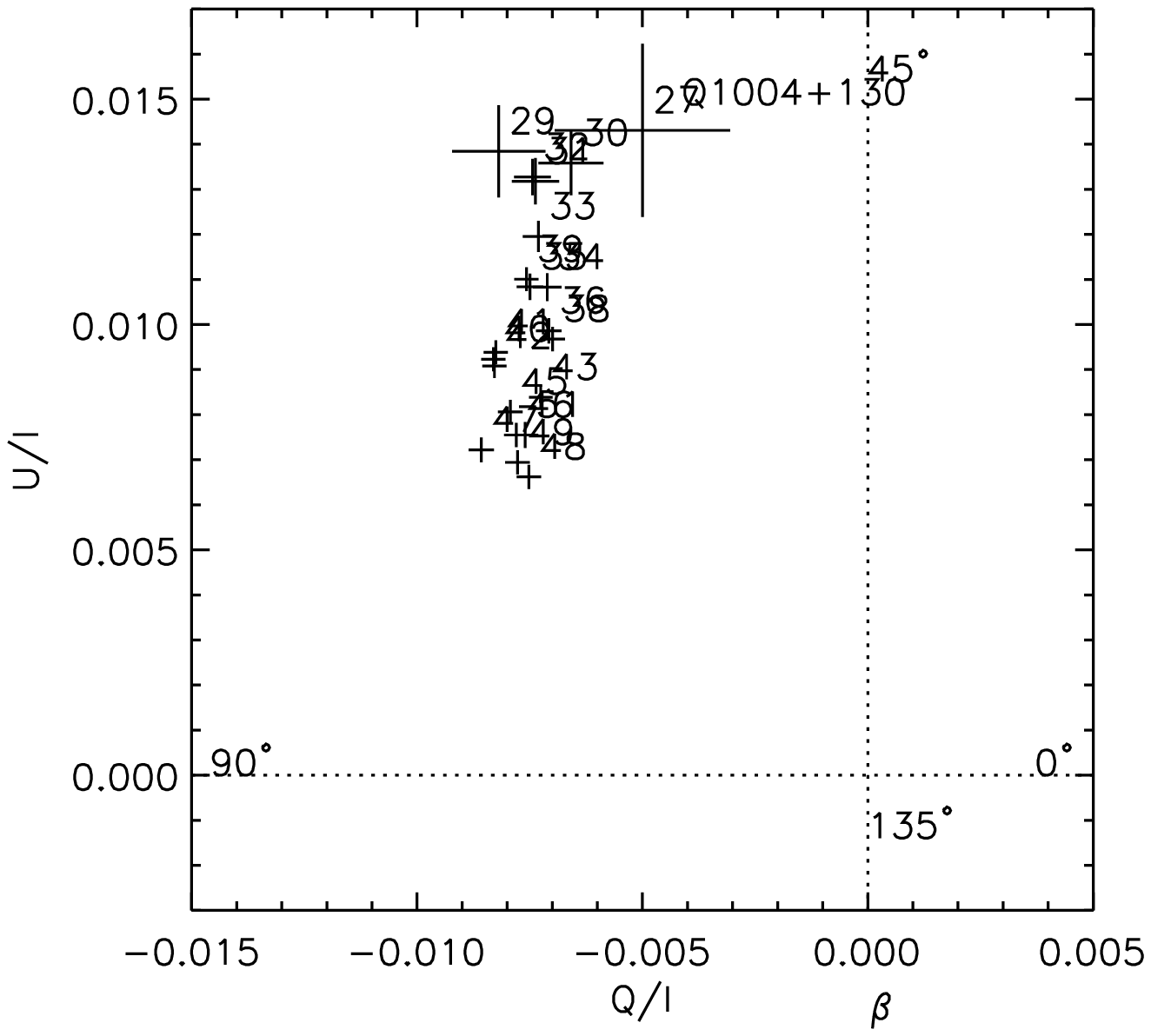}
 \caption{The same as Fig.\ref{res-4C09.72-qu}, but for Q1004+130.}
 \label{res-Q1004+130-qu}
\end{figure}

\subsection{Q1004+130} 

\subsubsection{ISP or second polarization component~$^1$}

The polarization gradually increases toward shorter wavelengths in
this object, and the PA rotates also gradually
(Fig.\ref{res-Q1004+130}). The data points on the $q$-$u$ plane align
linearly (Fig.\ref{res-Q1004+130-qu}), and its alignment probably
suggests that it has one polarization component at PA $\sim45$\degr\
and another component with $P$ being roughly wavelength-independent
and larger than $\sim0.8$\% based on the distance between the
alignment line and the origin (for this lower limit case, the PA of
this component is at PA $\sim90$\degr).  The latter component might be
an ISP: we do not have stars with large estimated distance within
$\sim 4$\degr, but at radii $4-5$\degr, several stars show a
polarization roughly aligned at PA 75-85\degr, though the polarization are
all low $\sim0.1$\% (one has an estimated distance of 100pc).  We need
a good observation of a Galactic star to check the ISP.

\subsubsection{Polarization and the BAL property}

In any case, the polarization property is quite different from the
previous objects. The polarization seems to consist of
two components: one of them is probably quite wavelength-independent,
but the other has quite a blue polarized flux, which might be due to
dust scattering. 

This quasar is one of the few Broad Absorption Line (BAL) quasars having
both a well-defined radio jet axis and good spectropolarimetric data.  The
evidence for the BAL property of this quasar was presented in
\citet{Wi99}.  The spectropolarimetry data of this quasar has been
documented by \citet{An96}, and our data are qualitatively consistent
with their results in terms of the polarization magnitude and PA
rotation, but their integrated polarization PA (though covering up to
longer wavelengths) is a little larger than ours (by $\sim 10$\degr), which
might be due to variability.  

The overall alignment between the polarization PA (67\degr) and radio
jet axis (121\degr) is intermediate. However, the component at
PA $\sim45$\degr\ is roughly perpendicular to the radio axis, and the
second component might be due to an ISP as described above.

In the other two cases of BAL quasars with FRII jet axis and optical
spectropolarimetric data (FIRST J101614.3+520916, \citealt{Gr00}; LBQS
1138-0126, \citealt{Br02}), the relations between the polarization PA
and jet axes are again intermediate and PA rotations are seen. For both
cases, $P$ is quite large and the ISP seems to be quite small based on
very small $\ebv$, suggesting that the PA rotation would originate
from two intrinsic polarization components rather than from an ISP
contamination.

\begin{figure}
 \includegraphics[width=80mm]{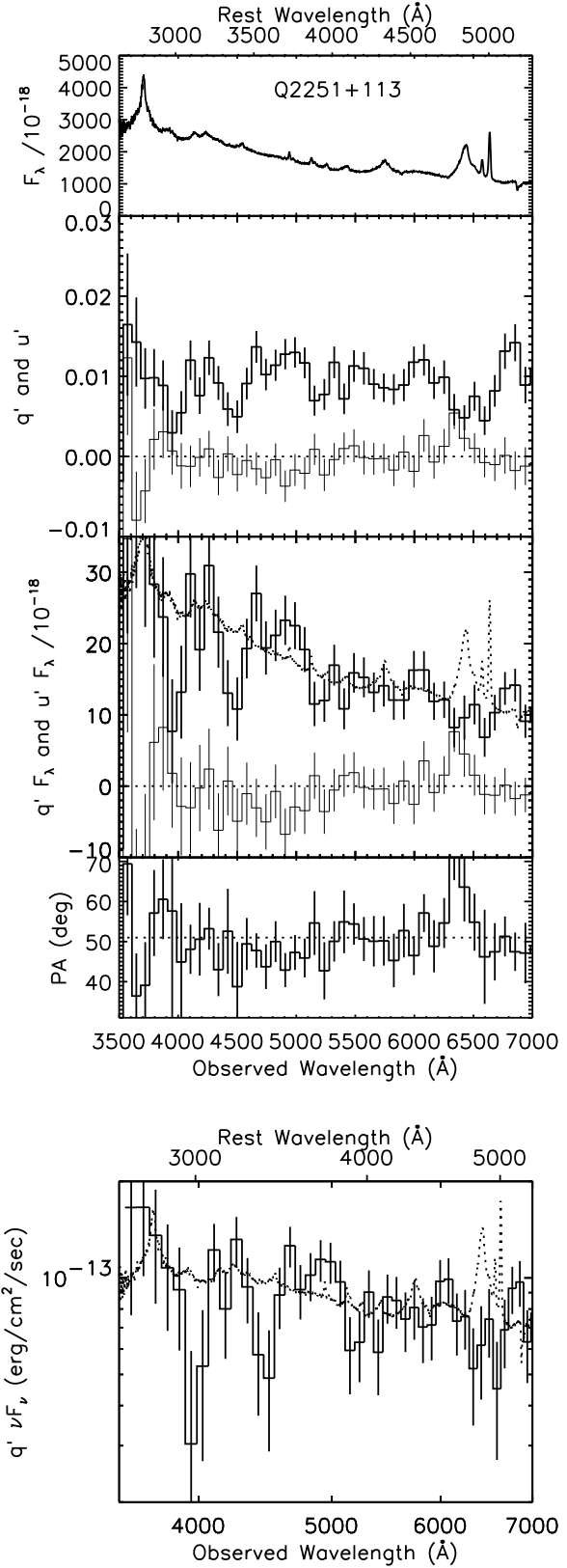}
 \caption{ The same as Fig.\ref{res-4C09.72}, but for Q2251+113.}
 \label{res-Q2251+113}
\end{figure}

\begin{figure}
 \includegraphics[width=80mm]{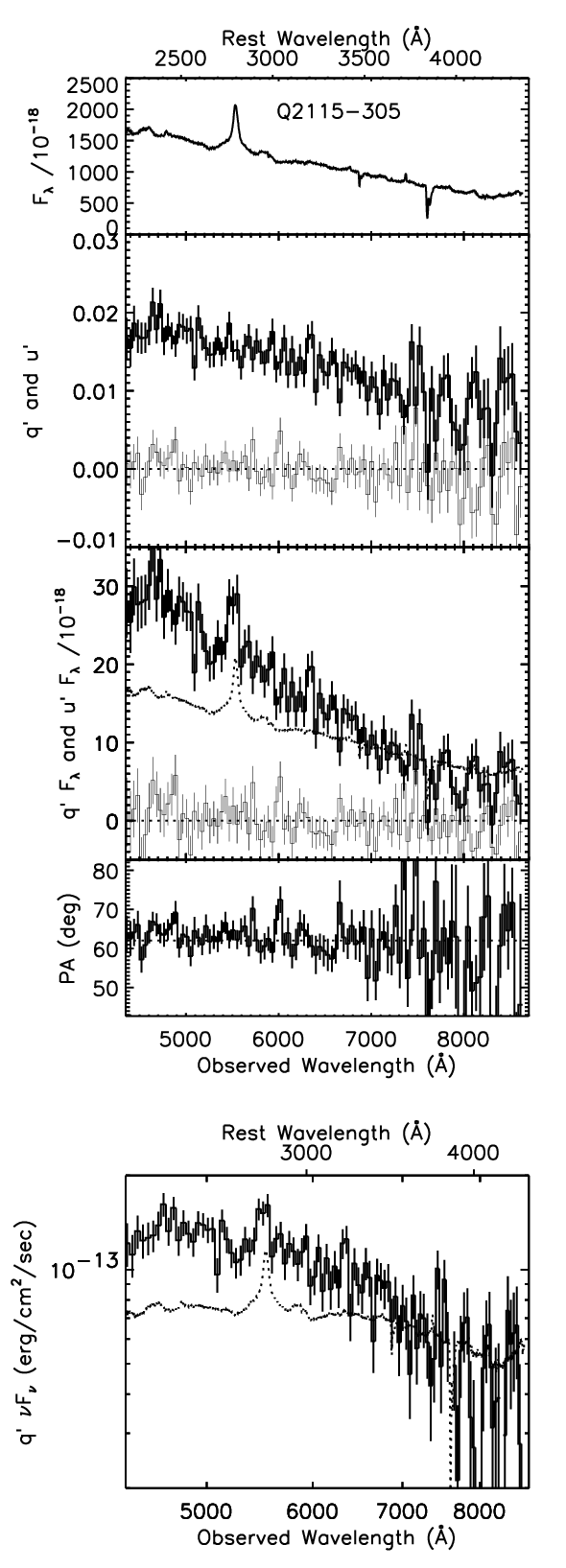}
 \caption{ The same as Fig.\ref{res-4C09.72}, but for Q2115-305. The
 reference axis of $q'$ and $u'$ is taken along the PA of the
 polarization integrated over the whole wavelength region, rather than
 at 4000-4731\AA.}
 \label{res-Q2115-305}
\end{figure}

\begin{figure}
 \includegraphics[width=80mm]{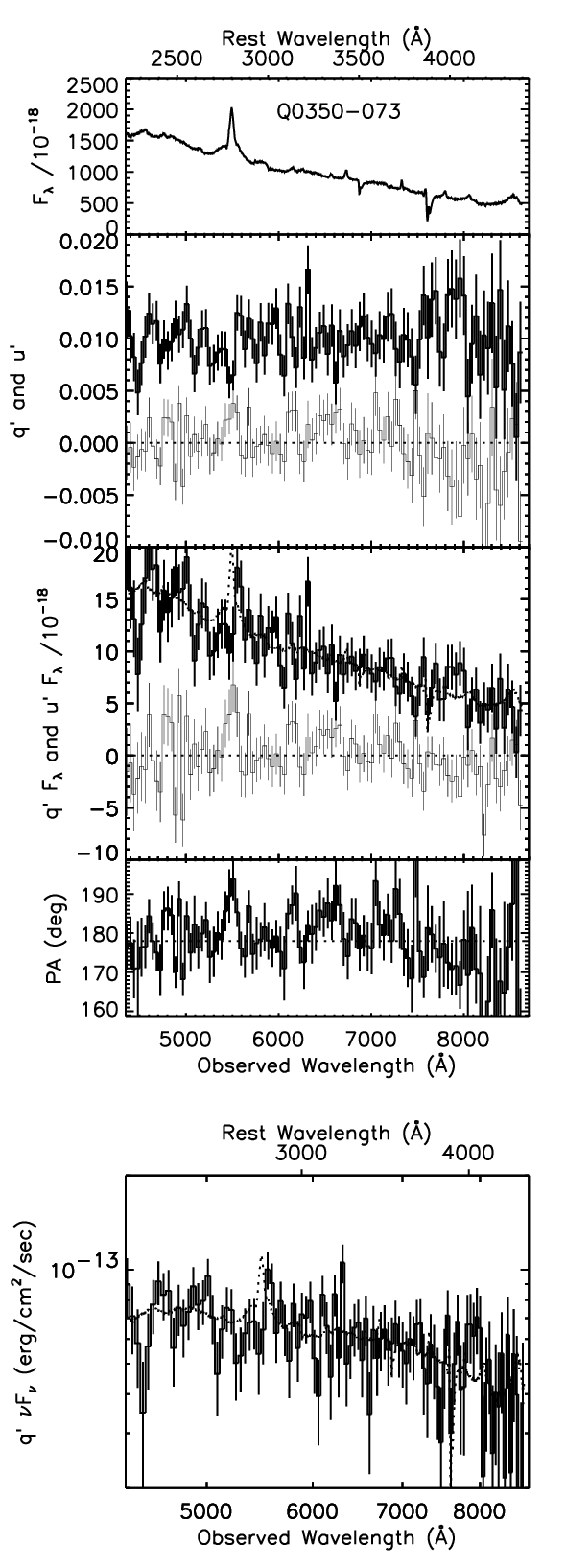}
 \caption{ The same as Fig.\ref{res-4C09.72}, but for Q0350-073. The
 reference axis of $q'$ and $u'$ is taken along the PA of the
 polarization integrated over the whole wavelength region, rather than
 at 4000-4731\AA.}
 \label{res-Q0350-073}
\end{figure}

\begin{figure}
 \includegraphics[width=80mm]{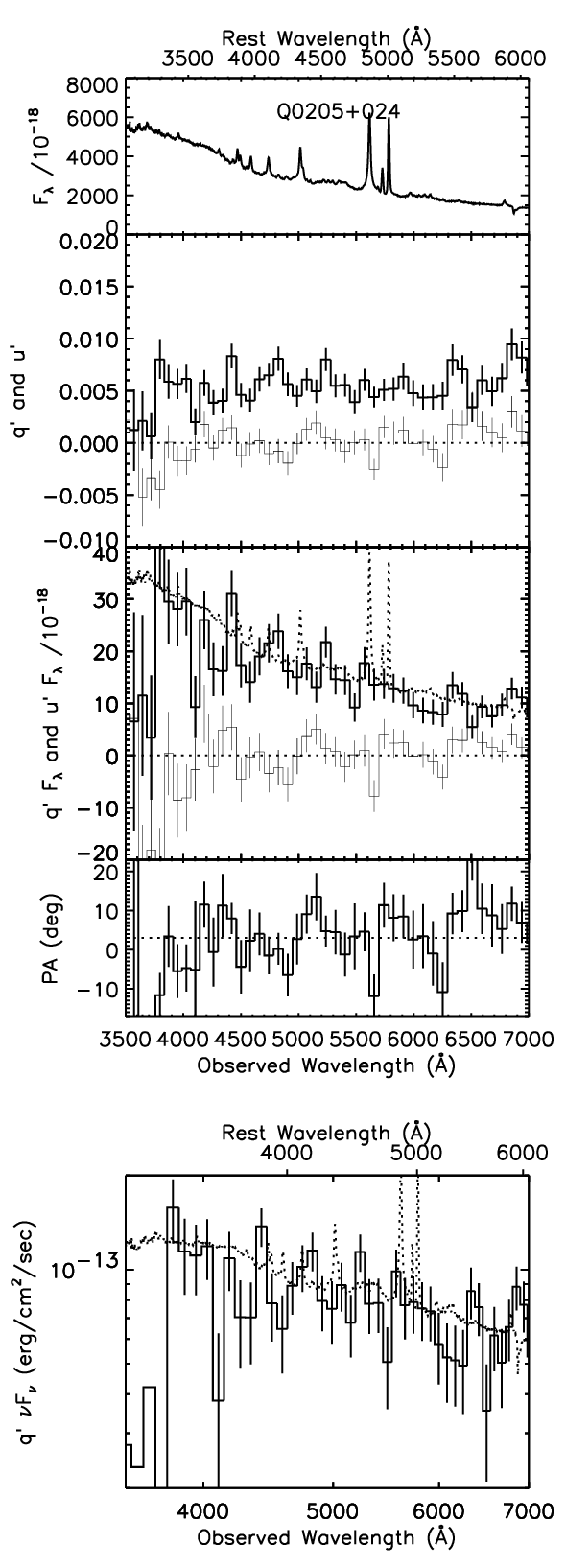}
 \caption{ The same as Fig.\ref{res-4C09.72}, but for Q0205+024.}
 \label{res-Q0205+024}
\end{figure}

\subsection{Q2251+113} 

The polarization looks constant over all the wavelengths, except at
the emission lines (Fig.\ref{res-Q2251+113}).  The blue side of the
H$\beta$ line seems to be polarized at a different PA from the
continuum.  However, there might be some effect from an ISP, since the
polarization measurements of several stars within 5\degr\ from the
quasar show an ISP at a consistent PA of 100-105\degr with $P= 0.2$\%.

\subsection{Q2115-305} 

Polarization is rising towards the bluer side, and thus the polarized
flux is bluer than the total flux (Fig.\ref{res-Q2115-305}).  The MgII
line is polarized in the same way as the neighboring continuum
(Table~\ref{tab-mg}). This may well be a case of scattering by dust
outside the BLR.  Several stars around the quasar (within 4\degr) show
a small polarization ($P=0.1-0.4$\%), and the PA tendency is not clear
although two of them with large estimated distances (2.4kpc and
0.5kpc) have PAs rather close to the polarization PA of Q2115-305.

\subsection{Q0350-073} 

The polarization looks almost constant, though slightly higher around
rest 4000\AA, and the polarized flux shape under the small blue
bump looks the same as the total flux (Fig.\ref{res-Q0350-073}).
There seems to be some PA rotation at the red end of the spectrum.
MgII line might be polarized at a different PA from the continuum. 
Only small polarizations for Galactic stars ($P<0.2$\%) are found
within 10\degr\ from the quasar, and no clear PA tendency is seen.

\subsection{Q0205+024}

This is a quasar which shows the spectrum of narrow-line Seyfert 1
galaxies.  The polarization level is small (Fig.\ref{res-Q0205+024}). It
looks constant over 4000\AA\ and the ratio is consistent with unity but
with a rather large uncertainty.  Two stars (within 5\degr\ from the
quasar) show 0.2-0.3\% polarization at 140-160\degr.

\begin{figure}
 \includegraphics[width=80mm]{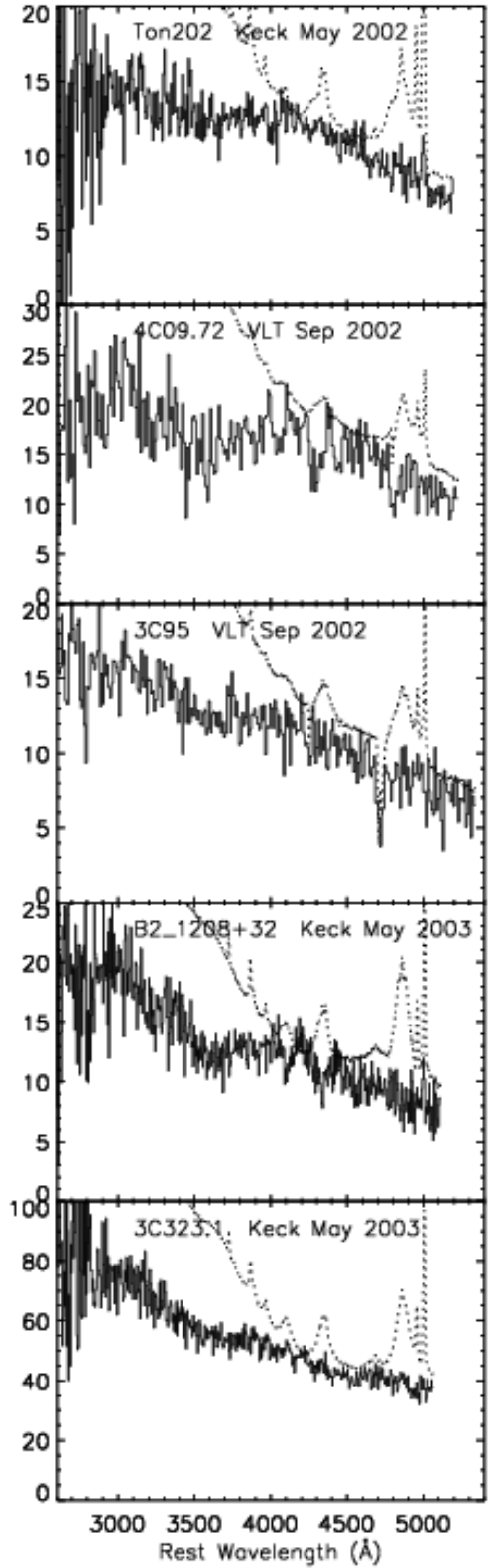} 

 \caption{Comparison of the polarized flux spectrum, or $q' F_{\lambda}$
 spectra, for Ton 202, 4C09.72, 3C95, B2 1208+32 and 3C323.1 (solid
 lines). The spectrum of Ton 202 is from the data in Paper I.  Flux has
 been corrected for Galactic reddening. For 4C09.72, 3C95 and 3C323.1,
 the spectrum is after the ISP correction, i.e., the same spectra as in
 Fig.\ref{res-4C09.72-cor100}, \ref{res-3C95-cor140} and
 \ref{res-3C323.1-cor090} are shown. The dotted line is the total flux
 spectrum scaled to approximately match the polarized flux spectrum at
 the red side.}

 \label{res-5obj}
\end{figure}

\begin{figure}
 \includegraphics[width=80mm]{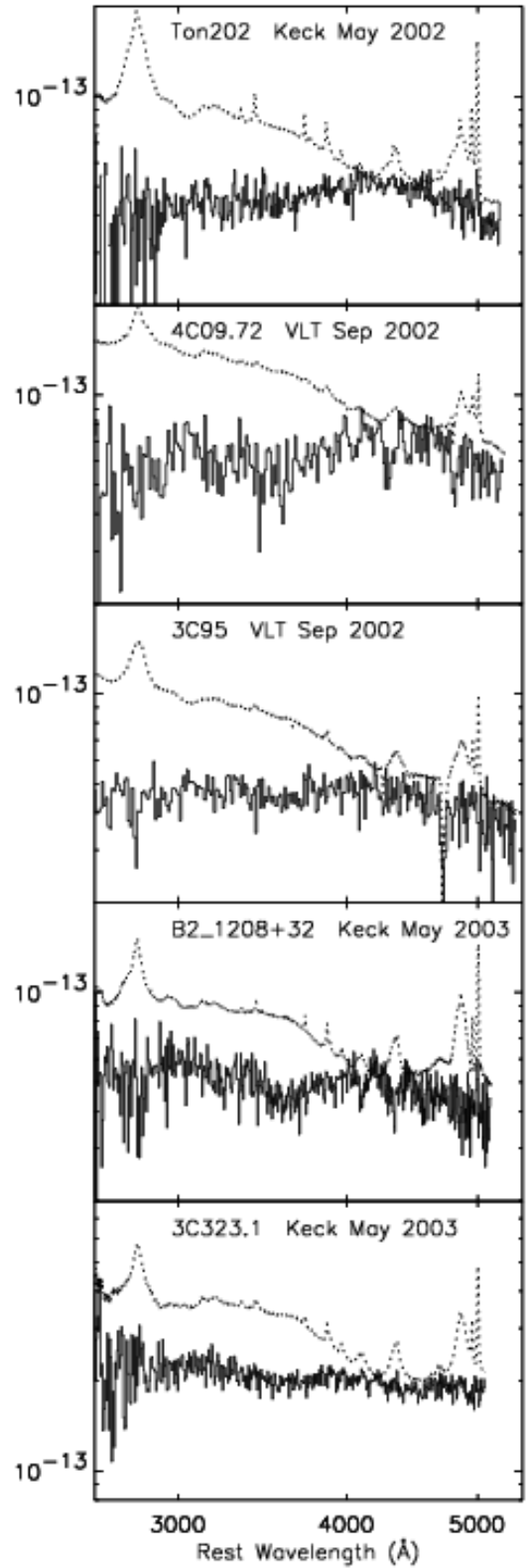}
 \caption{The same as Fig.\ref{res-5obj}, but in $\nu F_{\nu}$ with both
 axes in log scale.}
 \label{res-5obj-nFn}
\end{figure}

\section{Discussion}\label{sec-disc}

\subsection{The Balmer edge feature}

\subsubsection{Observed edge feature}

In Fig.\ref{res-5obj} we show the comparison of the $q' F_{\lambda}$
spectrum, i.e. essentially the polarized flux spectrum, for Ton 202
from Paper I and for 4C09.72, 3C95, 3C323.1 (after the ISP correction)
and B2 1208+32. The wavelengths are all in the rest frame of each
quasar.  The $q' \nu F_{\nu}$ spectra for all five objects are
compared in Fig.\ref{res-5obj-nFn}.  Note that in the ISP corrections,
the uncertainty in the PA of the ISP leads to some uncertainty in the
ISP-corrected polarized flux shape, but it is qualitatively unchanged
as we have described in the previous section (though the shape in
3C323.1 is a little more ambiguous than in 4C09.72 and 3C95).

The comparison shows that the slope of the polarized flux of all five
quasars longward of $\sim$4000\AA\ is essentially the same as that of
the total flux continuum, but the slope has a down-turn at
$\sim$4000\AA\ (i.e. a local maximum at $\sim4000$\AA; we note in one
case, 3C323.1, this seems to be at a substantially shorter wavelength
than in others), around where the small blue bump starts in the total
flux.  In addition, another change (an up-turn) of the slope is seen at
around 3600\AA\ (i.e. the spectra have local minima at $\sim3600$\AA).
This slope up-turn looks more evident in B2 1208+32, while both of the
slope changes look very weak in 3C323.1.  The slope up-turn was already
suggested in Paper I for Ton 202, but the comparison of these five
quasars rather strengthens the suggestion, since the position of this
slope up-turn looks roughly similar in the five polarized flux spectra.

Based on all these, we identify this spectral feature, which is
composed of these two slope changes, as a Balmer edge absorption
feature: (1) the slope change at $\sim 4000$\AA\ would be the start
(long wavelength limit) of a broadened Balmer edge absorption feature 
(2) the
change of the slope at $\sim$ 3600\AA\ would be the position of the
maximum opacity, with the flux shape at the shorter wavelength side
being formed by the combination of the intrinsic continuum color and
the $\nu^{-3}$ decrease of the continuum opacity, although the slope
looks still very red in the first three objects.  The feature may be
widespread, since the edge features in the five objects look rather
similar, and a few other objects show behavior which may also be
similar, but which is less clear.

In addition, there seem to be more spectral features in the shorter
UV wavelengths. We will briefly describe those in section
\ref{sec-disc-uv}.

\subsubsection{Simplest Interpretations}

The polarized flux of all these quasars essentially show no emission
lines, and thus the polarization is confined to the continuum.  These
quasars have lobe-dominant radio jets, and the direction of the optical
continuum polarization (E-vector direction) is in all cases parallel to
the radio structure, as we discuss in the next section.  The unpolarized
broad emission lines in our quasars lead to the inference that the
polarization originates {\it interior} to the BLR.\footnotemark[2] The
exact cause of the polarization is not well understood in these quasars,
but we have discussed some simple possibilities in Paper I, all based on
some form of electron scattering.  Dust scattering is unlikely, since
the region is thought to be interior to the BLR and thus within the dust
sublimation radius.  The electron scattering might originate in the
atmosphere of putative plane-parallel scattering-dominated
disk. However, this case suffers from the wrong prediction of the
polarization direction, i.e. perpendicular to the symmetry axis of the
disk which is thought to be along the jet direction (see Paper I for
more discussion).  Alternatively, the polarization is caused by an
optically thin oblate pure electron scattering region surrounding the
BBB emitter (with the BBB emission being unpolarized, e.g. completely
depolarized by Faraday depolarization with magnetic fields in the disc
atmosphere; \citealt{AB96,ABI98}).  This produces the right polarization
direction if the minor axis of the oblate shape is along the jet.  In
this latter case, the polarized flux is simply proportional to the BBB
emission.  In the case of electron scattering in the disk atmosphere,
the polarized flux is being produced as a part of the BBB emission and
is thus intrinsic to the BBB emission.  Therefore, in both of these
cases, the Balmer edge feature seen in our quasars should be intrinsic
to the BBB emission.

\footnotetext[2]{These properties are quite in contrast with Seyfert 2
galaxies where the broad emission lines and continuum are both polarized
in the same way, and the polarization is very high and perpendicular to
the radio structure \citep{An93}. Based on these properties, the
polarization in these type 2 objects is thought to be caused by polar
scattering of the radiation from a hidden nucleus and BLR along the jet
direction by matter in a region {\it outside} the BLR.}

To summarize, our most simple interpretation is that the observed
Balmer edge absorption feature is intrinsic to the BBB emission.  If
this is correct, then the observed feature directly indicates that the
BBB emission is indeed thermal, and its emitter is optically thick.
However, there is much to learn from the details of the polarized
flux features, which tend to repeat from object to object, if we
can identify them in theoretical models.
We also discuss other possibilities below in section
\ref{sec-disc-alt}.

\begin{figure}
 \includegraphics[width=80mm]{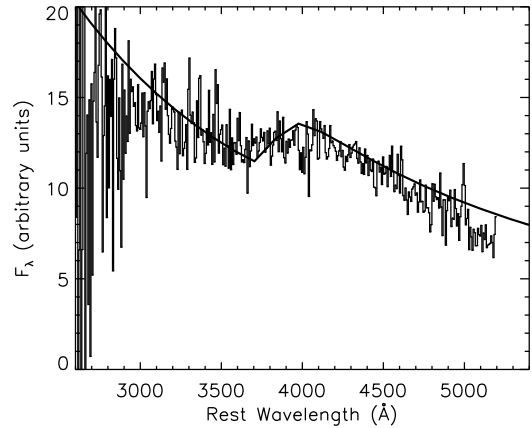}

 \caption{ As an illustration, a spectrum (thick smooth line) for a
 face-on disk model around a Kerr ($a/M=0.998$) black hole of mass
 $8\times10^9$~M$_\odot$, accreting at 2~M$_\odot$~yr$^{-1}$ is
 plotted over the May 2002 polarized flux spectrum of Ton202 (corrected
 for the Galactic reddening). The
 model spectrum has arbitrary normalization.}
 
 \label{model}
\end{figure}

\subsubsection{Broadening and a disk atmosphere model}

The onset of the edge feature looks broadened over the whole range 
from $\sim3600$\AA\ to
$\sim4000$\AA. This could be at least partly due to                      
the high-order Balmer absorption lines and possibly metal                
lines as well. However, the observed feature suggests                    
a broadening due to a high velocity dispersion in                         
the gravitational potential well.                                         
Very naively, in terms of a simple black-body disk, the expected          
velocity dispersion looks roughly consistent with the observed extent    
of broadening.  The radius $r(4000$\AA) emitting the radiation around     
4000\AA\ is given as                                                      
\begin{eqnarray*}
\lefteqn{ r(4000{\rm \AA}) /r_g } \\
&& \simeq 3.5 \times 10^2
\left( \frac{\eta}{0.1} \right)^{-1/3}
\left( \frac{L/L_{\rm Edd}}{0.1} \right)^{1/3}
\left( \frac{M}{10^8 M_{\sun}} \right)^{-1/3}, 
\end{eqnarray*}
where $M$, $\eta$ and $L_{\rm Edd}$ are the mass of the black hole,
radiative efficiency ($L=\eta \dot{M} c^2$ with accretion rate
$\dot{M}$) and Eddington luminosity, respectively, and $r_g = GM/c^2$.
The radius $350 r_g$ corresponds to $v \simeq 0.05c$ for a Keplerian
rotation, thus roughly consistent with the apparent broadening.

Detailed disk model spectra (Hubeny et al. 2000) are roughly in
agreement with these simple estimates, although considerable flux in
the Balmer edge region can arise from smaller radii in some disk
models.  As mentioned in section~\ref{sec-intro}, accretion disk
modeling of the Balmer edge region is complicated by the fact that the
relevant annuli contain internal hydrogen ionization zones.  These
result in very steep, convectively unstable temperature gradients, as
well as density inversions.  This only adds to the considerable
uncertainty in the vertical structure of such models.

As an illustration, Fig.\ref{model} shows a face-on disk model around a
Kerr ($a/M=0.998$) black hole of mass $8\times10^9$~M$_\odot$, accreting
at 2~M$_\odot$~yr$^{-1}$.  The model spectrum has arbitrary
normalization, and is plotted on top of the Ton~202 Keck data from May
2002.  Disks viewed at higher inclinations will have the edge smeared
out further and shifted toward the blue.  The model does not include any
lines, particularly high order Balmer lines and metal lines that may
help shift the edge to longer wavelengths.  Note that the model spectrum
is the total flux spectrum, and some polarizing mechanism would still
need to be invoked. (The intrinsic polarized flux produced by the disk
model at nonzero inclinations has a much stronger edge feature, and is
polarized in the wrong direction.)

If the Balmer edge arises from a disk, an atmosphere contributing to the 
edge at some radius might change on a time scale comparable to the thermal 
time at the effective photosphere.  This will be less than the local 
thermal time of the entire disk thickness at that radius, which is
\begin{eqnarray*}
\tau\sim0.2~{\rm yr}\left({\alpha\over0.1}\right)^{-1}
\left({r\over100 r_{\rm g}}\right)^{3/2}\left({M\over10^8{\rm M}_\odot}\right),
\end{eqnarray*}
where $\alpha$ is the Shakura \& Sunyaev (1973) viscosity parameter. 
This might be consistent with the time scale between observations. 
However, the disk would have to do this in a coordinated fashion at all 
radii producing flux in the Balmer edge region, and this probably requires 
reprocessing of radiation from the inner disk.  (Such reprocessing is 
already required in disk models in order to explain the near simultaneous 
optical/UV continuum variability observed in quasars.)

\begin{figure}
 \includegraphics[width=80mm]{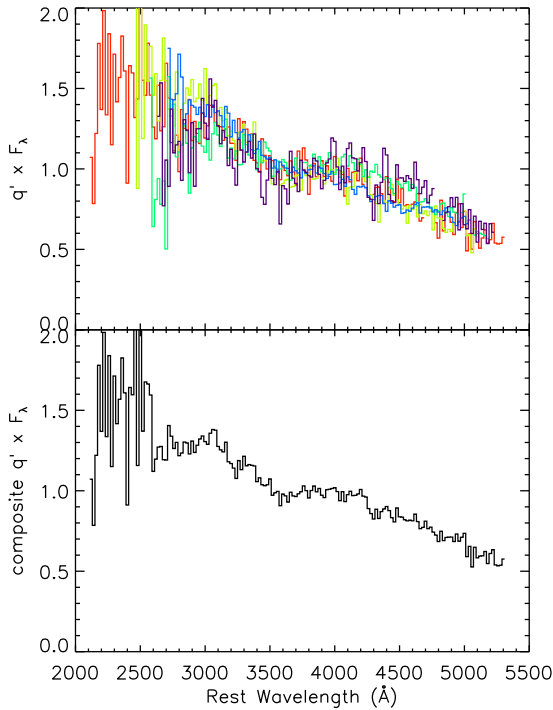}
 
 \caption{In the upper panel, the five polarized flux spectra shown in
    Fig.\ref{res-5obj}, normalized by the mean at 2800-5000\AA, are
    over-plotted with a larger wavelength bin of 20\AA.  In the lower
    panel the average of these five spectra with equal weighting is
    shown. Both panels are in $F_{\lambda}$.}
 \label{composite}
\end{figure}

\begin{figure}
 \includegraphics[width=80mm]{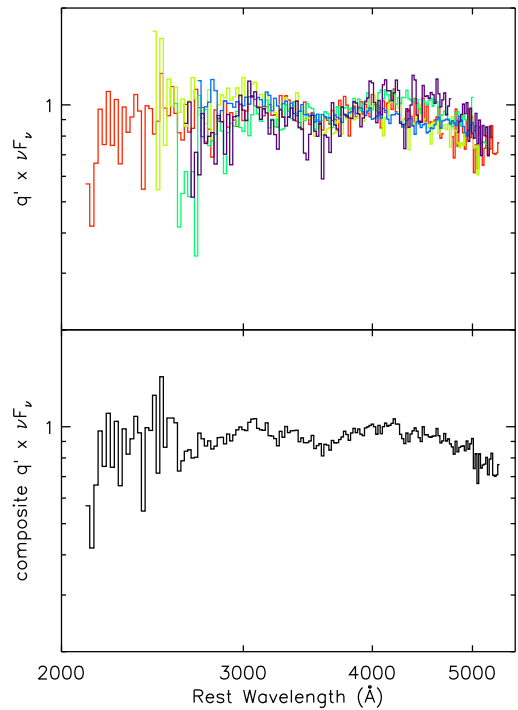}

 \caption{The same as Fig.\ref{composite}, but
    in $\nu F_{\nu}$ with both axes in log scale. } 
 \label{composite-nFn}
\end{figure}

\subsubsection{Alternative possibilities}\label{sec-disc-alt}

As discussed in Paper I, we cannot rule out some alternative
explanations.  One possible case would be that the feature is due to
foreground absorption.  However, a simple foreground absorption is
unlikely, due to the lack of a huge Lyman edge absorption in
quasars. This is expected in a simple foreground absorption case,
since the Lyman edge is a resonant feature, unlike the Balmer edge.
In particular, for the five objects shown in Fig.\ref{res-5obj}, the
ones with available far-UV spectra (4C09.72 and Ton202;
\citealt*{La93}) do not show such a strong Lyman edge absorption.

Alternatively, our Balmer edge absorption feature could be imprinted
in the scattering medium itself, i.e. both scattering and
absorption are occurring in the same region.  This is actually the favored
model for
the circumstellar disks of Be stars.  In these systems, the
radiation from the stellar photosphere is electron-scattered by the
circumstellar disk. Due to the absorption opacity in the disk medium
competing with electron scattering, absorption signatures are
imprinted in the scattered light spectrum. Thus a large Balmer edge is
seen in the polarized flux (e.g. \citealt*{Bj91}; note that their
Fig.1 shows $F_{\lambda}$ and $P$, not directly $P \times
F_{\lambda}$).

The same thing might be happening to our quasars. The edge feature
could be imprinted by the circumnuclear scattering region --- this
scattering region might possibly be the BLR clouds. However, our
argument is based on the empirical fact that there is essentially no
feature seen in the broad-line wavelengths, either in absorption or in
emission.  This fact (1) suggests that the spatial scale of the
scattering region is smaller than that of the BLR as we have argued,
(2) seems to be against any simple scattering model involving {\it in
situ} absorption.

Also, the fact that the edge feature looks quite broadened might
require quite specific conditions (e.g. high velocity dispersion) for
the medium producing the feature. If the feature is imprinted in the
scattering region, these conditions might put the scattering region 
as deep in the potential well as
a part of the continuum emitter: this is equivalent to one of our
simple pictures, i.e. the polarization and the feature in the
polarized flux originate from the BBB emitter and thus are intrinsic
to the BBB emission.

It is possible that the outer part of the disk itself scatters the
radiation from the inner regions. This would give the right polarization
direction parallel to the disk symmetry axis (which is supposed to be
along the radio jet axis).  The Balmer edge feature could be intrinsic
to the radiation from the inner regions or could be imprinted by the
outer part of the disk.  In either case, the medium producing the
feature must satisfy the constraints we discussed above: that the broad
lines are essentially unseen and the edge is broadened.

\subsection{UV spectral features}\label{sec-disc-uv}

We have so far concentrated on the spectral feature at the Balmer edge
region. However, there are apparently more spectral features in the
polarized flux at shorter wavelengths in at least some of the objects
shown in Fig.\ref{res-5obj}: a local peak at $\sim3050$\AA, and/or
absorption shortward of $\sim3050$\AA, and a possible absorption at
$\sim2600$\AA. These look clearer in the overplotted and
composite spectra shown in
Fig.\ref{composite} and \ref{composite-nFn}. To produce the composite,
we first normalized the five polarized flux spectra by the mean flux
between 2800\AA\ and 5000\AA\ in the rest frame of each quasar
excluding the wavelengths affected by telluric absorptions, binned the
spectra with a 20\AA\ width (shown in the upper panel), and then took
the average of these five with equal weighting (the lower panel).

The identification and interpretation of these possible spectral
features is not still clear, but they might be related to FeII
absorption and possibly the Bowen resonance-fluorescence lines which are
expected often to be highly polarized.  One of the authors (RA) would
offer US\$100,000 to anyone who successfully identifies them.  To be
decisive on these possible features, we need more data with a higher S/N
at this wavelength region on these and other quasars.

\begin{table*}
\begin{minipage}{180mm}

  \caption{Radio properties along with polarization properties. 
  Luminosities are calculated assuming $H_0=70$km/s/Mpc, $\Omega_{\rm m} = 0.3$ and 
  $\Omega_{\Lambda} = 0.7$. The values for $\nu L_{\nu}$ are from our spectroscopic
  measurements, so they are only approximate. $L_{\rm 5G}$ is the radio
 luminosity at rest-frame 5GHz, calculated assuming $f_{\nu} \propto \nu^{-0.5}$.
  PA$_{\rm opt}$ and PA$_{\rm cor}$ are the PA of the optical polarization at
 4000-4731\AA\ before and after the ISP correction, respectively (from Table~\ref{tab-pol}).
  Reference for 
  the radio map is: 1...\citet{Hu98}, 2...\citet{Re99}, 
  3...\citet{Pr93}, 4...\citet{Ke94}, 5...\citet{Ru84}, 6...\citet{An85},
  7...\citet{Fa77}.}

  \begin{tabular}{lccccccccccc}
  \hline
  Name & z & $\nu L_{\nu}$ (erg/s) & $L_{\rm 5G} $ & Radio        & PA$_{\rm opt}$ & PA$_{\rm cor}$ & $\Delta$PA & $r_P$ & EW ratio & $\log{R_V}$ & ref \\
       &   & at 4000\AA            & (W/Hz/sr)     & Axis (\degr) &   (\degr)         & (\degr)        & (\degr) \\
  \hline

  \multicolumn{8}{l}{Parallel} \\

  Q0414-060   & 0.775 & 46.07 & 25.77 & 152 & 162 &     & 10 & $0.88\pm0.12$ & $0.83\pm0.61$ & 1.78 & 2 \\ 
  3C95        & 0.616 & 46.01 & 25.79 & 166 &   1 &   9 & 23 & $0.61\pm0.01$ & $0.21\pm0.14$ & 1.13 & 3 \\ 
  4C09.72     & 0.433 & 45.89 & 25.06 & 143 & 119 & 131 & 12 & $0.54\pm0.01$ & $0.04\pm0.17$ & 1.44 & 4 \\ 
  Q0003+158   & 0.450 & 45.80 & 25.21 & 114 &  94 &  87 & 27 & $0.61\pm0.05$ & $0.36\pm0.24$ & 1.98 & 4 \\ 
  3C323.1     & 0.264 & 45.41 & 25.11 &  20 &  16 &  10 & 10 & $0.69\pm0.01$ & $0.08\pm0.04$ & 1.73 & 4 \\   
  B2 1208+32  & 0.388 & 45.35 & 24.65 &   3 &  29 &     & 26 & $0.72\pm0.01$ & $0.23\pm0.07$ & 1.45 & 7 \\
  Ton202      & 0.366 & 45.21 & 24.64 &  53 &  69 &     & 16 & $0.67\pm0.01$ & $0.20\pm0.08$ & 1.05 & 4 \\
  Q2349-014   & 0.174 & 44.96 & 24.60 & 170 & 155 &     & 15 & $0.58\pm0.11$ & $0.88\pm0.48$ & 1.69 & 6 \\ 

  \\
  \multicolumn{8}{l}{Intermediate} \\

  Q0405-123   & 0.573 & 46.67 & 26.04 &  13 & 139 & (160?) & 54 & $0.77\pm0.07$ & $2.49\pm1.41$ & 2.08 & 1 \\ 
  Q1004+130   & 0.240 & 45.49 & 24.72 & 121 &  67 & (45?)  & 54 & $1.22\pm0.02$ & $0.52\pm0.24$ & 0.51 & 4 \\

  \\
  \multicolumn{8}{l}{Perpendicular} \\

  Q2115-305   & 0.979 & 46.44 & 26.33 & 149 &  55 &     & 86 & $1.69\pm0.20$ & $1.27\pm0.26$ & $<$2.06 & 2 \\ 
  Q0350-073   & 0.962 & 46.34 & 26.32 &  97 & 170 &     & 73 & $0.98\pm0.10$ & $1.06\pm0.39$ & 0.96 & 2 \\ 
  Q2251+113   & 0.326 & 45.43 & 25.17 & 138 &  51 &     & 87 & $0.85\pm0.09$ & $1.50\pm0.33$ & 1.00 & 4 \\ 

  \\
  \multicolumn{8}{l}{Not classified} \\
  Q1912-550   & 0.402 & 45.26 & 24.72 &     &   5 &   1 &    & $0.45\pm0.02$ & $0.56\pm0.11$ &      &   \\
  Q0205+024   & 0.155 & 44.98 & 22.62 & 80? &   3 &     & 77?& $0.62\pm0.15$ & $1.24\pm0.61$ &      & 5 \\ 

  \hline
  \end{tabular}
  \label{tab-radio}
\end{minipage}
\end{table*}

\subsection{Radio properties}

In Table \ref{tab-radio}, we have summarized the radio properties of our
quasar sample along with the polarimetric properties.  The quasars
observed are all radio-loud except Q0205+024.  For these radio-loud
quasars, radio maps with resolved structures are available in the
literature except for Q1912-550. All of them show edge-brightened FR-II
morphology.

\begin{figure}
 \includegraphics[width=80mm]{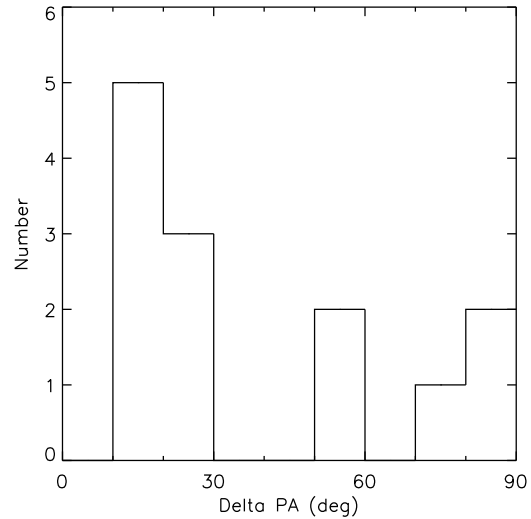}
 \caption{The histogram for the difference of the radio axis PA and optical polarization PA.}
 \label{hist}
\end{figure}

It is known that there is a strong statistical tendency for the
optical polarization E-vector direction to lie parallel to the radio
structural axis in quasars and Seyfert 1s (\citealt{SAM79, MS84,           
RS85,An83,An02}), though there are some cases where the optical polarization         
direction lies rather perpendicular to the radio axis (\citealt{MS84,      
An02, Sm02}).  Some quasars in our sample also show a perpendicular        
relationship (2 of them are included in Moore and Stockman sample), while  
most of the others show parallel polarization. In Fig.\ref{hist}, the
histogram of the difference of these two PAs ($\Delta$PA) is shown for
our sample. It is not certain whether this distribution indicates
bimodality (e.g. a dip test [\citealt{HH85}] does not reject
unimodality with a significant confidence level).  The bimodality is
also uncertain in the large samples such as those of \citet{MS84} and
\citet{RS85} (again, a dip test does not give a statistically
significant result).  In Table \ref{tab-radio}, we have classified the
quasars into three groups simply as 'parallel' ($\Delta$PA $\le$
30\degr), 'intermediate' (30\degr\ $< \Delta$PA $<$ 60\degr), and
'perpendicular' (60\degr\ $\le$ $\Delta$PA). Some properties seem to
have a possible $\Delta$PA dependence as we discuss below.

\subsubsection{$r_P$, line polarization and Balmer edge}

Most of the quasars in the parallel group show a decrease in
polarization shortward of 4000\AA, meaning an (at least partially)
unpolarized small blue bump.  We show the polarization ratio $r_P$ in
Table~\ref{tab-radio} (taken from Table~\ref{tab-pol}).  Those
in the parallel group also show essentially unpolarized emission lines:
many of them have a tight upper limit on the broad line EW in the
polarized flux compared to that in the total flux, as we show in
Table~\ref{tab-radio} the ratio of the former to the latter (for
H$\beta$, or MgII if H$\beta$ not available; see Tables \ref{tab-hbeta}
and \ref{tab-mg}).  It also seems that the quasars in the parallel group
tend to have the Balmer edge absorption feature in the polarized flux:
all the five quasars shown in Fig.\ref{res-5obj} are in this parallel
group, and the feature is possibly seen in another parallel quasar
Q0003+158.

\begin{figure}
 \includegraphics[width=80mm]{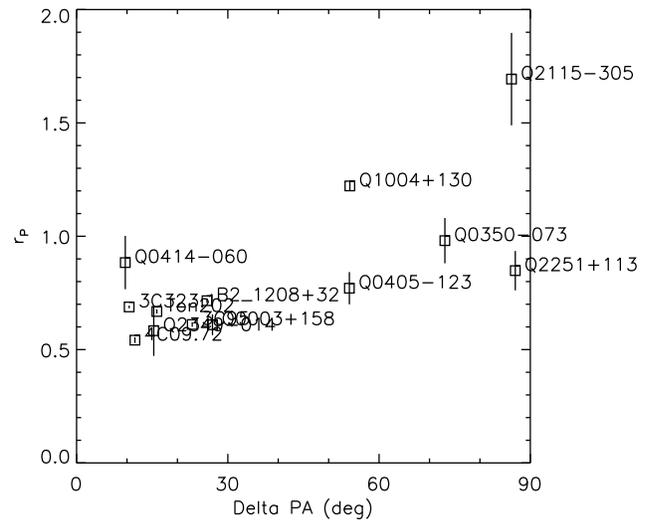}
 \caption{The ratio of the polarization at 2891-3600\AA\ to that at
 4000-4731\AA\ is plotted against $\Delta$PA.}
 \label{dpa-pratio}
\end{figure}

On the other hand, the quasars in the perpendicular group do not seem to
show the $P$ decrease shortward of 4000\AA\ ($r_P \sim 1$ or $>1$),
which indicates that they tend to have a polarized small blue
bump. While the constraint on $r_P$ is still not tight in two of the
three quasars, the other one (Q2115-305) shows $P$ increase instead of
decrease.  The broad lines also look polarized in these objects: this
is quite certain in Q2115-305; marginal in Q0350-073; in Q2251+113, the
line polarization is puzzling in that the PA rotation seen at the lines
is not present in the small blue bump wavelengths.  We have plotted
$r_P$ against $\Delta$PA in Fig.\ref{dpa-pratio}, which illustrate the
correlation between them (95.9\% statistical significance in the
Spearman's rank correlation coefficient, and 96.2\% in the Kendall's).

\begin{figure}
 \includegraphics[width=80mm]{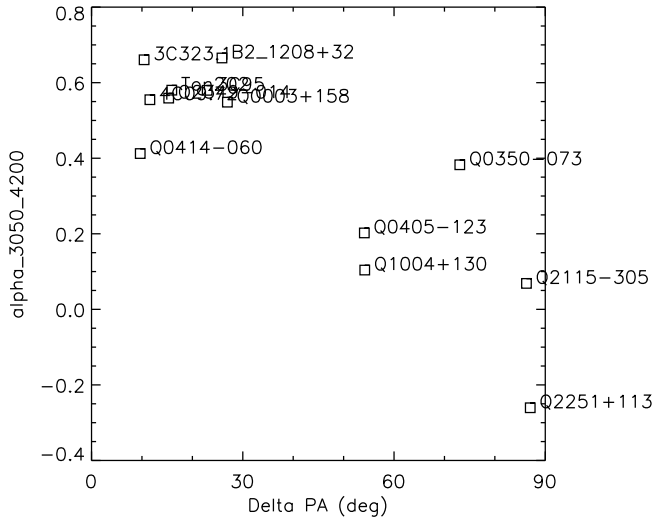}
 \caption{The spectral index $\alpha$ ($f_{\nu}\propto\nu^{\alpha}$)
 corresponding to the flux ratio at 3050\AA\ and 4200\AA\ is plotted
 against $\Delta$PA.}
 \label{dpa-alpha}
\end{figure}

Note also that $r_P$ depends on the strength of small blue bump in the
total flux. In fact, the small blue bump seems to be stronger in the
parallel ones and weaker in the perpendicular ones. This is shown in
Fig.\ref{dpa-alpha}.  As a rough strength measure for the small blue
bump, we simply take the ratio of the total flux at 3050\AA\ to that at
4200\AA\ (so of course this also depends on the color of the underlying
continuum), and convert it to the corresponding spectral index $\alpha$
where $f_{\nu}\propto\nu^{\alpha}$. This index is plotted against
$\Delta$PA, which illustrates the correlation (99.8\% statistical
significance in the Spearman's rank correlation coefficient, and 99.9\%
in the Kendall's).  Thus, $r_p$ is probably measuring the effects from
both the polarization and the strength of the small blue bump.

For the intermediate group, we note that Q0405-123, as we described in
the previous section, shows a polarization PA rotation shortward of
rest 4000\AA, which is similar to 4C09.72 (and 3C95 or 3C323.1) and
may be due to ISP contamination.  The PA of the intrinsic
polarization could be at $\sim$160\degr, which would put this object
close to the 'parallel' group ($\Delta$PA $\sim$ 33\degr).  Note also
that Q1004+130 may have an ISP contamination, and the intrinsic
polarization PA could be at $\sim$45\degr, which would put it into a
'perpendicular' group ($\Delta$PA $\sim$ 76\degr). Even before this
correction, this object looks similar to Q2115-305 in terms of the
rising $P$ toward shorter wavelengths.  These possible re-evaluations
of $\Delta$PA look consistent with the properties described above for
the parallel and perpendicular groups, though we did not detect
H$\beta$ line polarization in Q1004+130.

\subsubsection{$R_V$ and viewing angle}

For the ones with resolved radio structure, the ratio $R$ of the core to
lobe luminosity can provide a rough indication for the viewing angle,
the angle of our line of sight to the jet axis \citep{WB86}.  Or, a
better quantity is the ratio of the core luminosity to optical
luminosity \citep{WB95}, denoted as $R_V$.  We have calculated this
latter ratio $R_V$ for each quasar in our sample with available core
flux measurements, following \citet{WB95}. The values are shown in
Table~\ref{tab-radio}.  Note that the value for Q2115-305 is regarded as
an upper limit since its available map is coarse and no flat-spectrum
core seems to be well isolated.  The values for the two perpendicular
ones and Q1004+130 are smaller than the others, suggesting that the
perpendicular ones have a larger viewing angle than the parallel ones
(this tendency is also seen in the values of $R$).

\subsubsection{Possible interpretation for these and other AGN}

The overall tendencies for the small blue bump polarization, broad line
polarization, and viewing angles could be understood if we suppose that
(1) electron scattering is occuring in a flat or oblate compact
($\sim$100-1000$r_g$?) optically thin region interior to the BLR, and (2)
electron or dust scattering is occuring in an extended ($>$ pc)
optically thin polar region exterior to the BLR, as deduced from the
unified model.

When the viewing angle is small, the scattered light from the
$\sim$100-1000$r_g$ flat
compact region dominates, resulting in a small parallel polarization of
the continuum only.  The objects have Type 1 spectra.  When the viewing
angle is somewhat larger, and the the inner disk-like flat scattering region
becomes occulted or otherwise ineffective, the $>$pc-scale region
dominates: this would cause a small perpendicular polarization when the
BBB and BLR aren't occulted or just attenuated leading to a
perpendicular-polarized Type 1, and a Type 2 when these components do
become occulted.  In fact the continuum of the perpendicular ones in our
sample might be redder than the parallel ones, based on the flux ratio
at 3050\AA\ and 4200\AA\ which would depend both on the color of the
continuum under the small blue bump and the strength of the small blue
bump (the continuum color longward of the small blue bump wavelengths
could not be obtained reliably from our data for high-redshift ones in
particular).

For the polarization of Seyfert 1 galaxies, Smith et al. (2004) have
already suggested that two different scattering regions exist, one in an
inner equatorial region and the other in an outer polar region. The
difference from our case is that those Seyfert 1 galaxies with the
parallel property have polarized broad lines, and the polarization PA is
different from that of the continuum with the polarization magnitude
lower than that of the continuum (e.g. \citealt{Sm02}). This is also the
case in some broad-line radio galaxies (\citealt{Co99}).  In these
cases, the inferred inner equatorial scattering region would be
comparable in size to the BLR.

The distinction between the ones with essentially unpolarized lines
(i.e. most of the parallel quasars in our sample) and the ones with
partially polarized lines is not yet clear, but it might be related to optical
luminosity.  Many of the objects with partially polarized lines are of a
rather low optical luminosity ($\nu L_{\nu}$ at 4000\AA\ $\la 10^{45}$
erg/s), while the ones with unpolarized lines in our sample are of a
rather high optical luminosity ($\nu L_{\nu}$ at 4000\AA\ $\ga 10^{45}$
erg/s; see Table~\ref{tab-radio}).  However, an object like 3C382, which
has partially polarized broad lines, has a rather high optical
luminosity ($\nu L_{\nu}$ at 4000\AA\ $\sim 2 \times 10^{45}$ erg/s).
We need a larger sample for the distinction to be clarified.

\section{Conclusions}

In our previous paper, we have reported our detection of a Balmer edge
feature in one quasar, Ton202, which was observed in May 2002.  We have
since implemented VLT/FORS1 and Keck spectropolarimetry in Sep 2002 and
May 2003, respectively, on 14 more quasars with a known $\sim$1\%
optical polarization, and also re-observed Ton 202 in May 2003.  Some of
these observations are affected by interstellar polarization, but the
measurements have been corrected for this effect reasonably well.  Among
the newly observed 14 quasars, the high or relatively high S/N polarized
flux of at least four quasars (4C09.72, 3C95, 3C323.1, B2 1208+32) shows
essentially no emission lines.  We have found that these four quasars
show a Balmer edge feature in the polarized flux, which is quite similar
to that seen in Ton202 observed in May 2002.

Based on the unpolarized broad emission lines, the polarized flux is
thought to originate interior to the BLR, and the Balmer edge feature
is most simply interpreted as an intrinsic spectral feature of the Big
Blue Bump emission. In this case, the edge absorption feature
indicates that the BBB emission is thermal and optically thick.
However, we have also discussed other alternative interpretations.

The Balmer edge feature identified in the polarized flux of these five
quasars is entirely a new finding. The feature is apparently widespread,
based on its similarity in these well observed objects, and the
possible similar behavior of the polarized flux at around 4000\AA\ in
two other objects (Q0003+158, Q1912-550). There also seem to be more
spectral features in the shorter rest UV wavelengths of the polarized
flux, which are common to multiple objects.

In the re-observation of Ton202 in May 2003, we did not find a similar
dramatic feature. This is apparently due to real polarization
variability. This is not unexpected since the polarization is thought
to originate from a small scale, less than the size scale of broad
line region. The polarization variability of normal quasars (aside
from synchrotron dominated objects) has been reported in the
literature. This object, as well as other objects should be followed
up to confirm the variability.

Among the 15 quasars in our sample, 13 are radio-loud quasars with an
available radio map.  While many of them show the optical polarization
roughly parallel to the radio structural axis, some of them show the
polarization rather perpendicular to the radio axis (however, the
bimodality is not certain).  Although the sample is too small to draw
strong conclusions, the parallel ones tend to show an unpolarized and
stronger small blue bump (thus shows a polarization drop shortward of
4000\AA), unpolarized broad lines, a Balmer edge absorption feature in
the polarized flux, and a smaller viewing angle, while the
perpendicular ones tend to show a polarized and/or weaker small blue
bump, polarized broad lines, and a larger viewing angle.

Based on our spectropolarimetric data and other data in the
literature, the parallel type 1 objects seem to have at least
partially unpolarized broad lines.  The ones with (almost completely)
unpolarized lines might have a higher optical luminosity than the ones
with partially unpolarized lines. However, a much larger sample is
required to clarify this distinction.

\section*{Acknowledgments}

This research has made use of the NASA/IPAC Extragalactic Database (NED)
which is operated by the Jet Propulsion Laboratory, California Institute
of Technology, under contract with the National Aeronautics and Space
Administration. This research has also made use of the SIMBAD database,
operated at CDS, Strasbourg, France. The authors thank A. Barth for
providing calibration data. The work by RA was supported in part by NSF
grant AST-0098719. The work by OB was supported in part by 
NASA grant NAG5-13228.


\label{lastpage}

\end{document}

%% file: tab_pol.tex
\begin{tabular}{lcrrrrc}
  \hline
  Name & z & \multicolumn{2}{c}{2891-3600\AA} & \multicolumn{2}{c}{4000-4731\AA} & $r_P$ \\
  \cline{3-4} \cline{5-6}
       &   & P (\%) & PA (\degr)              & P (\%) & PA (\degr)              & \\
  \hline
              Q0003+158 & 0.450 & $0.72\pm0.02$ & $ 95.0\pm  0.8$ & $0.93\pm0.02$ & $ 94.4\pm  0.7$ & $0.77\pm0.03$ \\ 
\  ISP 103\degr, 0.45\% &       & $0.32\pm0.02$ & $ 83.8\pm  1.7$ & $0.52\pm0.02$ & $ 87.0\pm  1.3$ & $0.61\pm0.05$ \\ 
              Q0205+024 & 0.155 & $0.38\pm0.09$ & $166.8\pm  6.3$ & $0.60\pm0.03$ & $  3.0\pm  1.6$ & $0.62\pm0.15$ \\ 
                   3C95 & 0.616 & $0.75\pm0.01$ & $176.0\pm  0.3$ & $1.17\pm0.02$ & $  1.4\pm  0.4$ & $0.64\pm0.01$ \\ 
\  ISP 140\degr, 0.32\% &       & $0.71\pm0.01$ & $  8.6\pm  0.3$ & $1.17\pm0.02$ & $  8.9\pm  0.4$ & $0.61\pm0.01$ \\ 
              Q0350-073 & 0.962 & $1.01\pm0.04$ & $  0.9\pm  1.0$ & $1.03\pm0.10$ & $170.0\pm  2.7$ & $0.98\pm0.10$ \\ 
              Q0405-123 & 0.573 & $0.32\pm0.02$ & $129.9\pm  1.8$ & $0.42\pm0.03$ & $138.9\pm  1.8$ & $0.77\pm0.07$ \\ 
              Q0414-060 & 0.775 & $0.80\pm0.05$ & $163.2\pm  1.8$ & $0.90\pm0.10$ & $161.7\pm  3.2$ & $0.88\pm0.12$ \\ 
              Q1004+130 & 0.240 & $1.43\pm0.02$ & $ 59.9\pm  0.4$ & $1.17\pm0.01$ & $ 66.8\pm  0.3$ & $1.22\pm0.02$ \\ 
             B2 1208+32 & 0.388 & $1.01\pm0.01$ & $ 34.2\pm  0.3$ & $1.41\pm0.01$ & $ 28.9\pm  0.2$ & $0.72\pm0.01$ \\ 
                 Ton202 & 0.366 & $1.25\pm0.01$ & $ 66.9\pm  0.3$ & $1.87\pm0.02$ & $ 68.9\pm  0.2$ & $0.67\pm0.01$ \\ 
       Ton202 (May2002) & 0.366 & $1.19\pm0.02$ & $179.5\pm0.4^*$ & $2.11\pm0.01$ & $  0.2\pm0.2^*$ & $0.56\pm0.01$ \\ 
                3C323.1 & 0.264 & $0.78\pm0.01$ & $ 21.1\pm  0.4$ & $1.37\pm0.01$ & $ 15.8\pm  0.2$ & $0.57\pm0.01$ \\ 
\  ISP  90\degr, 0.90\% &       & $1.51\pm0.01$ & $ 10.2\pm  0.2$ & $2.19\pm0.01$ & $  9.6\pm  0.1$ & $0.69\pm0.01$ \\ 
              Q1912-550 & 0.402 & $1.14\pm0.02$ & $  6.3\pm  0.5$ & $1.75\pm0.02$ & $  4.5\pm  0.4$ & $0.65\pm0.01$ \\ 
\  ISP  10\degr, 0.70\% &       & $0.49\pm0.02$ & $  1.1\pm  1.3$ & $1.09\pm0.02$ & $  1.0\pm  0.6$ & $0.45\pm0.02$ \\ 
              Q2115-305 & 0.979 & $1.32\pm0.03$ & $ 62.0\pm  0.7$ & $0.78\pm0.09$ & $ 55.3\pm  3.3$ & $1.69\pm0.20$ \\ 
              Q2251+113 & 0.326 & $0.83\pm0.07$ & $ 49.7\pm  2.5$ & $0.98\pm0.06$ & $ 51.0\pm  1.7$ & $0.85\pm0.09$ \\ 
                4C09.72 & 0.433 & $1.02\pm0.01$ & $113.0\pm  0.3$ & $1.33\pm0.01$ & $119.2\pm  0.3$ & $0.77\pm0.01$ \\ 
\  ISP 100\degr, 0.70\% &       & $0.50\pm0.01$ & $131.4\pm  0.6$ & $0.93\pm0.01$ & $131.4\pm  0.4$ & $0.54\pm0.01$ \\ 
              Q2349-014 & 0.174 & $0.52\pm0.10$ & $144.5\pm  5.2$ & $0.89\pm0.05$ & $154.7\pm  1.5$ & $0.58\pm0.11$ \\ 
  \hline
\end{tabular}

%% file: tab_hb.tex
\begin{tabular}{lcccccccccc}
  \hline
  Name & \multicolumn{2}{c}{continuum region} & \multicolumn{4}{c}{H$\beta$ line region (4731-4934\AA)} & \multicolumn{4}{c}{OIII5007 line region (4982-5032\AA)} \\
  \cline{4-7} \cline{8-11}
       &  \multicolumn{2}{c}{(4537-4731, 5032-5185\AA)} & \multicolumn{2}{c}{polarization} & \multicolumn{2}{c}{EW (\AA)}& \multicolumn{2}{c}{polarization} & \multicolumn{2}{c}{EW (\AA)}  \\
  \cline{2-3} \cline{4-5} \cline{6-7} \cline{8-9} \cline{10-11}
       & P (\%) & PA (\degr)           & P (\%) & PA (\degr)      & $Q'$     & $I$    & P (\%) & PA (\degr)      & $Q'$     & $I$    \\
  \hline
              Q0003+158 & $0.92\pm0.04$ & $ 95.2\pm  1.3$ & $0.53\pm0.13$ & $ 94.\pm    7.$ & $ 54.\pm13.$ &  92. & $0.40\pm0.23$ & $ 79.\pm   17.$ & $ 14.\pm 8.$ &  31. \\ 
\  ISP 103\degr, 0.45\% & $0.52\pm0.04$ & $ 88.7\pm  2.2$ & $0.18\pm0.13$ & $ 69.\pm   20.$ & $ 33.\pm22.$ &      & $0.34\pm0.23$ & $ 42.\pm   19.$ & $ 21.\pm14.$ &      \\ 
              Q0205+024 & $0.60\pm0.05$ & $  4.1\pm  2.5$ & $0.75\pm0.37$ & $132.\pm   14.$ & $ 51.\pm25.$ &  42. & $0.56\pm0.35$ & $ 46.\pm   18.$ & $ 23.\pm14.$ &  25. \\ 
                   3C95 & $1.16\pm0.03$ & $  2.0\pm  0.8$ & $0.11\pm0.17$ & $170.\pm   42.$ & $  5.\pm 7.$ &  50. & $0.48\pm0.24$ & $ 54.\pm   15.$ & $  8.\pm 4.$ &  20. \\ 
\  ISP 140\degr, 0.32\% & $1.17\pm0.03$ & $  9.1\pm  0.8$ & $0.25\pm0.16$ & $ 40.\pm   19.$ & $ 11.\pm 7.$ &      & $0.75\pm0.24$ & $ 54.\pm   10.$ & $ 13.\pm 4.$ &      \\ 
              Q0405-123 & $0.45\pm0.05$ & $138.3\pm  3.3$ & $1.12\pm0.63$ & $ 96.\pm   16.$ & $ 45.\pm25.$ &  18. & $0.59\pm0.48$ & $ 55.\pm   23.$ & $ 22.\pm18.$ &  18. \\ 
              Q1004+130 & $1.11\pm0.02$ & $ 68.1\pm  0.4$ & $0.58\pm0.26$ & $134.\pm   13.$ & $  9.\pm 4.$ &  17. & $0.56\pm0.32$ & $ 71.\pm   16.$ & $  4.\pm 2.$ &   8. \\ 
             B2 1208+32 & $1.29\pm0.02$ & $ 28.1\pm  0.5$ & $0.29\pm0.08$ & $171.\pm    8.$ & $ 17.\pm 5.$ &  72. & $0.29\pm0.12$ & $ 36.\pm   12.$ & $  7.\pm 3.$ &  28. \\ 
                 Ton202 & $1.79\pm0.03$ & $ 69.5\pm  0.5$ & $0.33\pm0.14$ & $ 56.\pm   12.$ & $ 11.\pm 5.$ &  55. & $0.50\pm0.12$ & $ 79.\pm    7.$ & $ 13.\pm 3.$ &  40. \\ 
       Ton202 (May2002) & $2.07\pm0.02$ & $  0.3\pm0.3^*$ & $0.18\pm0.09$ & $  3.\pm 14.^*$ & $  6.\pm 3.$ &  73. & $0.46\pm0.08$ & $  8.\pm  5.^*$ & $ 11.\pm 2.$ &  45. \\ 
                3C323.1 & $1.42\pm0.02$ & $ 15.4\pm  0.3$ & $0.91\pm0.08$ & $ 95.\pm    3.$ & $ 33.\pm 3.$ &  51. & $1.19\pm0.13$ & $ 86.\pm    3.$ & $ 16.\pm 2.$ &  18. \\ 
\  ISP  90\degr, 0.90\% & $2.24\pm0.02$ & $  9.5\pm  0.2$ & $0.19\pm0.08$ & $134.\pm   13.$ & $  4.\pm 2.$ &      & $0.33\pm0.13$ & $ 78.\pm   11.$ & $  3.\pm 1.$ &      \\ 
              Q1912-550 & $1.89\pm0.04$ & $  5.1\pm  0.6$ & $1.18\pm0.14$ & $175.\pm    3.$ & $ 46.\pm 5.$ &  73. & $1.15\pm0.16$ & $176.\pm    4.$ & $ 24.\pm 3.$ &  39. \\ 
\  ISP  10\degr, 0.70\% & $1.23\pm0.04$ & $  2.4\pm  0.9$ & $0.70\pm0.14$ & $160.\pm    6.$ & $ 41.\pm 8.$ &      & $0.65\pm0.16$ & $161.\pm    7.$ & $ 21.\pm 5.$ &      \\ 
              Q2251+113 & $1.16\pm0.09$ & $ 50.1\pm  2.3$ & $1.75\pm0.39$ & $116.\pm    6.$ & $108.\pm24.$ &  72. & $1.13\pm0.63$ & $141.\pm   17.$ & $ 21.\pm12.$ &  23. \\ 
                4C09.72 & $1.32\pm0.02$ & $120.8\pm  0.4$ & $0.56\pm0.16$ & $101.\pm    8.$ & $ 14.\pm 4.$ &  33. & $0.65\pm0.24$ & $101.\pm   10.$ & $  7.\pm 3.$ &  13. \\ 
\  ISP 100\degr, 0.70\% & $0.96\pm0.02$ & $132.5\pm  0.6$ & $0.04\pm0.16$ & $  1.\pm   90.$ & $  1.\pm 6.$ &      & $0.02\pm0.24$ & $ 98.\pm   90.$ & $  0.\pm 3.$ &      \\ 
              Q2349-014 & $0.91\pm0.07$ & $159.7\pm  2.3$ & $0.80\pm0.43$ & $102.\pm   15.$ & $ 45.\pm24.$ &  51. & $0.28\pm0.58$ & $ 42.\pm   59.$ & $  7.\pm14.$ &  21. \\ 
  \hline
\end{tabular}

%% file: tab_mg.tex
\begin{tabular}{lcccccc}
  \hline
  Name & \multicolumn{2}{c}{continuum region} & \multicolumn{2}{c}{line region} & \multicolumn{2}{c}{EW (\AA)} \\
  \cline{2-3} \cline{4-5} \cline{6-7}
       & P (\%) & PA (\degr)           & P (\%) & PA (\degr)      & $Q'$ & $I$ \\
  \hline
              Q0003+158 & $0.60\pm0.06$ & $ 98.5\pm   2.7$ & $0.65\pm0.26$ & $104.\pm   11.$ & $ 41.\pm16.$ &  37. \\
\  ISP 103\degr, 0.45\% & $0.20\pm0.06$ & $ 89.2\pm   8.1$ & $0.27\pm0.26$ & $100.\pm   27.$ & $ 57.\pm54.$ &      \\
                   3C95 & $0.69\pm0.02$ & $170.2\pm   0.8$ & $0.05\pm0.13$ & $ 82.\pm   78.$ & $  2.\pm 6.$ &  33. \\
\  ISP 140\degr, 0.32\% & $0.60\pm0.02$ & $  3.4\pm   0.9$ & $0.35\pm0.12$ & $ 55.\pm   10.$ & $ 19.\pm 7.$ &      \\
              Q0350-073 & $0.97\pm0.06$ & $178.4\pm   1.8$ & $1.03\pm0.38$ & $ 33.\pm   11.$ & $ 33.\pm12.$ &  32. \\
              Q0414-060 & $0.70\pm0.10$ & $159.8\pm   3.9$ & $0.58\pm0.43$ & $134.\pm   22.$ & $ 36.\pm26.$ &  43. \\
             B2 1208+32 & $1.00\pm0.04$ & $ 34.8\pm   1.1$ & $0.61\pm0.16$ & $136.\pm    8.$ & $ 23.\pm 6.$ &  38. \\
                 Ton202 & $1.10\pm0.04$ & $ 66.4\pm   1.1$ & $0.20\pm0.13$ & $165.\pm   19.$ & $ 11.\pm 7.$ &  60. \\
       Ton202 (May2002) & $0.80\pm0.09$ & $175.0\pm 3.3^*$ & $0.45\pm0.16$ & $ 11.\pm 10.^*$ & $ 44.\pm16.$ &  78. \\
                3C323.1 & $0.82\pm0.05$ & $ 18.3\pm   1.8$ & $1.05\pm0.26$ & $ 79.\pm    7.$ & $ 46.\pm11.$ &  37. \\
\  ISP  90\degr, 0.90\% & $1.52\pm0.05$ & $  9.4\pm   1.0$ & $0.50\pm0.26$ & $ 61.\pm   15.$ & $ 12.\pm 6.$ &      \\
              Q1912-550 & $0.98\pm0.08$ & $  6.2\pm   2.4$ & $1.20\pm0.22$ & $  5.\pm    5.$ & $ 73.\pm13.$ &  55. \\
\  ISP  10\degr, 0.70\% & $0.36\pm0.08$ & $179.3\pm   6.5$ & $0.56\pm0.22$ & $177.\pm   12.$ & $ 97.\pm39.$ &      \\
              Q2115-305 & $1.52\pm0.05$ & $ 62.5\pm   1.0$ & $1.92\pm0.39$ & $ 72.\pm    6.$ & $ 34.\pm 7.$ &  27. \\
              Q2251+113 & $1.08\pm0.47$ & $ 69.5\pm  12.3$ & $6.30\pm1.71$ & $ 19.\pm    8.$ & $188.\pm51.$ &  35. \\
                4C09.72 & $0.94\pm0.03$ & $111.3\pm   1.0$ & $0.51\pm0.20$ & $105.\pm   11.$ & $ 13.\pm 5.$ &  25. \\
\  ISP 100\degr, 0.70\% & $0.40\pm0.03$ & $132.4\pm   2.3$ & $0.21\pm0.20$ & $175.\pm   27.$ & $ 13.\pm12.$ &      \\
  \hline
\end{tabular}